\documentclass{article}
\usepackage{graphicx} 
\usepackage{jheppub} 
\usepackage{hyperref}
\usepackage{amssymb}

\usepackage{amsthm}
\usepackage{amstext}
\usepackage{amsmath}
\usepackage{amsfonts}
\usepackage[linesnumbered,ruled,noline]{algorithm2e}
\usepackage{breqn}
\setkeys{breqn}{breakdepth={0}}
\usepackage{booktabs}
\usepackage{bbm}
\usepackage[german,english]{babel}
\usepackage{comment}
\usepackage{epstopdf}
\usepackage{multirow}
\usepackage{tikz}
\usepackage{verbatim}
\usepackage{subfig}

\usepackage{listings}
\usetikzlibrary{shapes,arrows}
\usetikzlibrary{decorations.pathmorphing}
\tikzstyle{startstop} = [rectangle,rounded corners=0.5cm, minimum width=2.5cm,minimum height=1cm,text centered, draw=black]
\tikzstyle{process} = [rectangle,rounded corners=0.2cm,minimum width=3cm,minimum height=1cm,text centered,text width=3cm,draw=black]
\tikzstyle{io} = [rectangle,minimum width=3cm,minimum height=1cm,text centered,text width =3cm,draw=black]
\tikzstyle{plaintext} = [rectangle,text centered,text width=3cm,minimum width=1cm,minimum height=0.5cm]
\tikzstyle{decision} = [trapezium, trapezium left angle = 70,trapezium right angle=110,minimum width=1cm,minimum height=0.5cm,text centered,text width=3cm,draw=black]
\tikzstyle{arrow} = [thick,->,>=stealth]

\newcommand\diff{{\rm d}}
\newcommand\B{{\mathcal{B}}}
\newcommand\Bp{{\mathcal{B}^\prime}}
\newcommand\C{{\mathcal{C}}}
\newcommand\F{{\mathcal{F}}}
\newcommand\FI{{I_{\alpha_1,\cdots,\alpha_n}}}

\newcommand\HT{{\mathcal{H}}}
\newcommand\K{{\mathcal{K}}}
\newcommand\R{{\mathcal{R}}}

\newcommand\I{{\mathcal{I}}}

\newcommand\Ord{{\mathcal{O}}}
\newcommand\Z{{{\mathcal{Z}}}}
\newcommand\Zm{{{\check{\mathcal{Z}}}}}
\newcommand\vl{{\vec{l}}}

\newcommand\idint{{\int_{\text{id}}}}

\newcommand\vp{{\vec{p}}}
\newcommand\vpnorm{{|\vp\,|}}
\newcommand\vz{{\vec{z}}}
\newcommand\vznorm{{|\vz\,|}}
\newcommand\bD{{\bar{D}}}
\newcommand\Res{{\text{Res}}}
\newcommand\Sec{{\text{Sec}}}
\newcommand\Sign{{\text{Sign}}}
\newcommand\Subsets{{\text{Subsets}}}
\newcommand\udth{{\underline{3}}}
\newcommand\udon{{\underline{1}}}
\newcommand\uot{{\underline{1}2}}

\newcommand\cut{\text{-cut}}
\newcommand\mir{\text{-mir}}
\newcommand\Mir{\text{-Mir}}
\newcommand\zp{{0^+}}

\newcommand\spcref{\cite{Britto:2004nc,Larsen:2015ped,Ita:2015tya,Boehm:2020ijp,Georgoudis:2016wff,Bern:1994cg,Berger:2008sj,Bern:1997sc,Bern:2004cz,Bern:2000dn,Bern:1994zx,Zhang:2016kfo} }

\newcommand\MMAC{{\small{\color{red}[MMA CHECK THIS EQUATION]}}}
\definecolor{myblue}{RGB}{20,70,160}

\theoremstyle{remark}
\newtheorem{remark}{Remark}

\def\ZW#1{\textcolor{orange}{{\bf\tt [ZW: #1]}}}

\def\Cut#1{\{#1\}\cut}
\def\mirror#1{\{#1\}\mir}

\definecolor{dark gray}{RGB}{100,100,255}

\preprint{TTP26-018, P3H-26-041}

\title{On the spanning cuts consistency problem in the IBP reductions of Feynman integrals}
\author[a,b]{Zihao Wu,}
\author[c]{Yingxuan Xu}
\affiliation[a]{State Key Laboratory of Nuclear Physics and Technology, Institute of Quantum Matter, South China Normal University, Guangzhou 510006, China}
\affiliation[b]{Guangdong Basic Research Center of Excellence for Structure and Fundamental Interactions of Matter, Guangdong Provincial Key Laboratory of Nuclear Science, Guangzhou 510006, China}
\affiliation[c]{Institute for Theoretical Particle Physics, Karlsruhe Institute of Technology (KIT), Wolfgang-Gaede-Straße~1, 76131 Karlsruhe, Germany}

\emailAdd{wuzihao@mail.ustc.edu.cn}
\emailAdd{yingxuan.xu@kit.edu}

\abstract{
The spanning cuts method is a powerful approach to reduce the cost of IBP reduction while computing Feynman integrals. However, its usage is limited due to the so-called consistency problem. It was unclear why the IBP reduction coefficients can be inconsistent with each other between different cuts. In this paper, we report a mechanism behind this inconsistency. We found that the IBP relations can be violated under the cuts, if we blindly erase the hidden terms that are proportional to the ``vanishing'' Feynman prescription parameters in the relations. In some cases, the cut introduces pinch singularities, which cancel the vanishing Feynman prescription parameters, making the hidden terms finite. In various cases, the error comes from omitting such finite hidden terms. We also claimed that the pinch singularity under the cuts are related to some hidden relations between the propagators. In this paper, we provide an algorithm and its implementation to find the linear hidden relations.
}

\begin{document}

\maketitle

\section{Introduction}

The evaluation of multi-loop Feynman integrals is one of the most important and challenging tasks in quantum field theory. In its modern workflow, the integration-by-parts (IBP) reduction~\cite{Tkachov:1981wb, Chetyrkin:1981qh} is often a bottleneck step in many frontier problems. In recent decades, various methods and software have been developed to perform IBP reduction~\cite{Baikov:1996iu,
Anastasiou:2004vj,Smirnov:2005ky,Smirnov_2006,Smirnov:2006tz,Smirnov:2006vt,
Smirnov:2006wh,Lee:2008tj,
Smirnov:2008iw,
Studerus:2009ye,Gluza:2010ws,Grozin:2011mt,Schabinger:2011dz,Smirnov:2012gma,vonManteuffel:2012np,
Lee:2013mka,Smirnov:2013dia,Lee:2014tja,Smirnov:2014hma,vonManteuffel:2014ixa,Ita:2015tya,Larsen:2015ped,Peraro:2016wsq,Bitoun:2017nre,Bohm:2017qme,
Maierhofer:2017gsa,Bohm:2018bdy,
Kosower:2018obg,Liu:2018dmc,Maierhofer:2018gpa,
Mastrolia:2018uzb,Bendle:2019csk,Frellesvig:2019kgj,Frellesvig:2019uqt,Guan:2019bcx,Klappert:2019emp,Maierhofer:2019goc, Peraro:2019svx,Smirnov:2019qkx,Frellesvig:2020qot,Klappert:2020aqs,Klappert:2020nbg,vonManteuffel:2020vjv,Liu:2021wks,Magerya:2022hvj,Usovitsch:2022wvr,Smirnov:2023yhb,Wu:2023upw,Guan:2024byi,Lange:2024mmz,Smirnov:2024onl,Lange:2025fba,Song:2025pwy,vonHippel:2025okr,Wu:2025aeg,Zeng:2025xbh,Shih:2026jfe,Huang:2024nij,Smirnov:2025prc,Huang:2026xnq}. Still, the reduction is difficult for integrals with large number of loops, external legs, or mass parameters.

In the effort of finding better algorithms for IBP reduction, the \textit{generalized unitarity cuts} (GUC) \spcref is proved to be useful. By setting some of the propagators on-shell, the integral families are usually simplified since plenty of its sectors become zero sectors. Based on the conjuncture that GUC keeps the IBP relation unchanged, since it does not affect the integrand and the boundary, the so-called \textit{spanning cuts} method \cite{Larsen_2016,Britto:2024mna} was introduced\footnote{See \cite{Bohm:2018bdy,Bendle:2019csk,Boehm:2020zig,Bern:2024adl,Chicherin:2025mvc} for its applications.} to simplify the IBP reduction. With a set of cuts that covers all the non-zero sectors of the family, a huge reduction problem can be decomposed to plenty of much simpler ones. As reported in \cite{Chicherin:2025mvc}, for example, the spanning cuts method is vital for breaking through the computational bottlenecks of nowadays's frontier problems. This method was also implemented in many public software \cite{Guan:2024byi,Wu:2025aeg,Smirnov:2025prc,Lange:2025fba,Crisanti:2026rbc}.

Despite its importance, the spanning cuts method itself suffers from the so-called \textit{consistency problem}: As what was believed, the IBP reduction coefficients of an integral should agree with each other under different cuts, inferred by the conjecture that IBP relations are unchanged under GUC. This is a requirement called the \textit{consistency condition}. However, many studies \cite{Guan:2024byi,Wu:2025aeg,Crisanti:2026rbc} have reported violations of this condition.

Since its discovery, people are researching the cause of this problem, as well as finding solutions to it. In Ref.~\cite{Guan:2024byi,Crisanti:2026rbc}, the consistency problem was reported to be having correlation to the so-called \textit{magic relations} \cite{Maierhofer:2018gpa,Smirnov:2013dia,Lee:2010ik}, which is a kind of IBP relation induced from a super sector but contains integrals only in its sub sectors. Ref. \cite{Guan:2024byi} has also designed a variation of the spanning cuts method to avoid the influence of magic relations.

Even with these studies, the cause of this problem is still unrevealed. We still need to know what exactly happened in the cases that the consistency condition breaks. It is still  mysterious why we see some magic relations breaks under the cuts, knowing that they can be expressed as linear combination of ordinary IBP relations, which we believe to be kept under the cuts.

In this paper, we will provide our answer to the question. We claim that the conjecture that \textit{IBP relations is unchanged under the cut} is not one-hundred-percent correct if we drop out the \textit{hidden terms} in the IBP relations. The hidden terms are terms proportional to the vanishing Feynman prescriptions. We are used to erase the hidden terms in the IBP relations since we believe that they vanish. However, in this paper, we will show that, under certain choice of cuts, they do not vanish. In these cases, the vanishing prescription parameters are canceled by pinch singularities introduced by the cuts, making the hidden terms in fact finite. If we blindly drop them out, the IBP relations will appear to be violated. We will show with an example that, if we keep those hidden terms, the IBP relation still holds under these cuts.

This paper is organized as follows. In Section \ref{sec: review}, we review some concepts. In Section \ref{sec:simple example}, we provide a simple but typical example, the one-loop massive bubble, to show how hidden terms spoils the IBP relations under the cut. We also raise a concept named \textit{hidden relations} to help us detect the possible pinch singularities under the cut. In Section \ref{sec: search linear hr}, we provide an algorithm and its implementation to find linear hidden relations. In Section \ref{sec: 2-loop special example}, we provide a more complicated two-loop example without linear hidden relation, where inconsistencies still occur. In Section \ref{sec: conclusion}, we summarize this paper.

\section{Review of concepts and methods}\label{sec: review}
\subsection{Feynman integrals and sectors}
A generic Feynman integral with $L$ loops is defined as,
\begin{equation}
I_{\alpha_1,\cdots,\alpha_n}=\int \prod_{j=1}^L \frac{\diff^D l_j}{i \pi^{D/2}} \, \frac{1}{\prod_{i=1}^n P_i^{\alpha_i}},
\end{equation}
where $\alpha_i \in \mathbb{Z}$ are integer indices, and the propagators $P_i$ are linear or quadratic polynomials in the loop momenta $l_j$ and external momenta $p_i$, given by
\begin{gather}
P_i = \sum_{j ,k=1}^L A_i^{jk} \, l_j \cdot l_k + \sum_{j=1}^L B_i^j \cdot l_j + E_i.
\end{gather}
Here, the coefficients $A_i^{jk}$ and $B_i^j$ depend on the external kinematics, and $E_i$ is a scalar function of external momenta. One should keep in mind that there are Feynman prescription parameters in the denominators, which we did not explicitly write down here.

The sector of an integral is defined to label which of the propagators appears in the denominator of the integrand. Exlicitly,

\begin{equation}
    \Sec(I_{\alpha_1,\cdots,\alpha_n})=\{i|\alpha_i>0\}
\end{equation}
\subsection{IBPs and spanning cuts}
The integration-by-parts (IBP) identities~\cite{Tkachov:1981wb,Chetyrkin:1981qh} are
\begin{gather}
0 = \int \prod_{j=1}^L \frac{\diff^D l_j}{i \pi^{D/2}} \, 
\frac{\partial}{\partial l_k^\mu} 
\left( \frac{v_k^\mu}{\prod_{i=1}^n P_i^{\alpha_i}} \right),
\label{IBP-relation}
\end{gather}
where $v_k^\mu$ is a vector linear in loop or external momenta. 

The Feynman integrals under a certain cut is defined as

\begin{gather}
I_{\alpha_1,\cdots,\alpha_n}|_{\Cut{\C}} = \int \prod_{j=1}^L \frac{\diff^D l_j}{i \pi^{D/2}} \,
\frac{1}{\prod_{i \notin \mathcal{C}} P_i^{\alpha_i}} \,
\prod_{k \in \mathcal{C}} \delta(P_k).
\end{gather}

if $\alpha_k=1$ for $k\in\C$. Where $\C$ is a sub set of $\{1,2,\cdots,n\}$. If $\alpha_k\neq1$, it is clearer to define GUC in the Baikov representation \cite{Baikov:1996cd,Baikov:1996rk,Baikov:2005nv}, where an integral is expressed as
\begin{equation}\label{eq:Baikov rep}
    I_{\alpha_1,\cdots,\alpha_n}=J\int_\Omega\diff z_1 \cdots \diff z_n P^\gamma \frac{1}{z_1^{\alpha_1} \cdots z_n^{\alpha_n}},
\end{equation}
where $J$ and $\gamma$ are constants. $P$ is a polynomial of $z_i$'s, and $\Omega$ is a boundary where $P=0$. The integrals under cut are
\begin{equation}\label{eq: C-cut}
    I_{\alpha_1,\cdots,\alpha_n}|_{\C\text{-cut}}:=J\int_{\Omega_\C}\prod_{i\notin\C} \diff z_i\prod_{i\in\C}\oint_0 \diff z_i\frac{P^\gamma}{{z_1}^{\alpha_1}\cdots {z_n}^{\alpha_n}},
\end{equation}

With this definition, we see that 
\begin{equation}
    \C\not{\subseteq}\Sec(\FI)\Rightarrow \FI|_{\Cut{\C}}=0
\end{equation}

Thus, we can produce plenty of zero-sectors using GUC, such that all the integrals belonging to the sector vanish. For some reduction problems, this significantly reduces the cost of the computation.

Bases on this, the \textit{spanning cuts} idea is as follows. Firstly, we find out a set of cuts
\begin{equation}
    \{\C_i\},
\end{equation}
such that for any non-zero sector $S$ in a family, 
\begin{equation}
    \{\C_i\}\cap \Subsets(S)\neq\{\}.
\end{equation}
Then, we perform IBP reductions for a certain integral $I$ on the spanning cuts, which should be faster, to get
\begin{equation}
    I|_{\Cut{\C_i}}=\sum_j c_j^i I_j|_{\Cut{\C_i}},
\end{equation}
where the R.H.S is the linear combination of master integrals under the cuts.

We then use the above results to reconstruct the IBP reduction result of the original problem, without cuts. We label it as
\begin{equation}
    I=\sum_j c_j I_j.
\end{equation}
If we admit the conjecture that IBP relations are unchanged under cuts, we have
\begin{equation}\label{eq:spc merge}
    c_j^i=\begin{cases}
        c_j,\, &\text{if } \C_i\subseteq\Sec(I_j)\\
        0,\, &\text{if } \C_i\not\subseteq\Sec(I_j)
    \end{cases}.
\end{equation}
We can use the above to reconstruct $c_j$. We pick up one $\C_i$ such that $\C_i\subseteq\Sec(I_j)$, then we have 
\begin{equation}
    c_j=c_j^i.
\end{equation}
When there are multiple choices, we require that all the corresponding $c_j^i$ of these choices should be the same. Otherwise, \eqref{eq:spc merge} does not hold. This is called the \textit{consistency condition} of the spanning cuts method. However, as stated, this condition is violated in many observations. This leads to the consistency problem studied in this paper.

\section{A simple example with inconsistency}\label{sec:simple example}
In this section, we provide one simple but typical example where the inconsistency occurs. 
\subsection{Example description}

\begin{figure}[hbtp]
\centering
\begin{tikzpicture}[scale=0.8]

\draw[line width=3pt](0,0) arc (180:-180:1);
\draw (-1,0)--(0,0);
\draw (2,0)--(3,0);

\node[font=\large\bfseries] at (1,1.45){$m^2$};
\node[font=\large\bfseries] at (1,-1.45){$m^2$};
\node[font=\large\bfseries] at (-1.3,0){$p$};
\node[font=\large\bfseries] at (3.3,0){$p$};

\end{tikzpicture}
\caption{A one-loop two-point massive bubble example}
\label{fig:imbubble}
\end{figure}
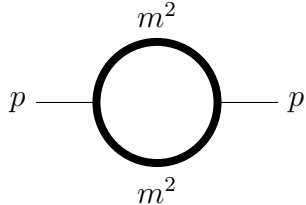

We consider the one-loop bubble with massive propagators and massless external legs shown in Figure \ref{fig:imbubble}. The propagators are defined as
\begin{equation}
    P_1=l^2-m^2,\, P_2=(l+p)^2-m^2.
\end{equation}
The external kinematic condition is
\begin{equation}
    p^2=0.
\end{equation}

The integrals in this family is then
\begin{equation}
    I_{\alpha_1,\alpha_2}=\int \frac{\diff^{D}l}{i \pi^{\frac{D}{2}}}\frac{1}{P_1^{\alpha_1}P_2^{\alpha_2}}.
\end{equation}

We pick up one of the IBP relations as
\begin{equation}\label{eq: IBP example im bubble}
        0=\int \frac{\diff^{D}l}{i \pi^{\frac{D}{2}}}\frac{\partial}{\partial l^{\mu}}\frac{p^\mu}{P_1P_2}
        =(I_{0,2}-I_{2,0})
\end{equation}
If we apply $\{1\}\cut$ on the equations, we have
\begin{equation}
    I_{0,2}|_{\{1\}\cut}=0
\end{equation}
If we admit the conjecture that GUC keeps the IBP relation \eqref{eq: IBP example im bubble} unchanged, we will have
\begin{equation}
    I_{2,0}|_{\{1\}\cut}=0
\end{equation}
However, this is \textbf{not true}. One can directly compute it as 

\begin{equation}\label{eq:1-cut spec calculation}
\begin{aligned}
    I_{2,0}|_{\{1\}\cut}&=-2\pi i\int \frac{\diff^{D}l}{i \pi^{\frac{D}{2}}}\delta^\prime((l^0)^2-|\vec{l}|^2-m^2)=2\pi i\int \frac{\diff^{D-1}\vec{l}}{i \pi^{\frac{D}{2}}}\int\diff\big( \frac{\diff l^0}{\diff P_1}\big)\delta((l^0)^2-|\vec{l}|^2-m^2)\\
    &=\int \frac{\diff^{D-1}\vec{l}}{i \pi^{\frac{D}{2}}}\int\frac{-2\pi i}{2 (l^0)^2}\delta((l^0)^2-|\vec{l}|^2-m^2)\diff l^0=\int \frac{\diff^{D-1}\vec{l}}{i \pi^{\frac{D}{2}}}\frac{-2\pi i}{2(|\vec{l}|^2+m^2)^{\frac{3}{2}}}\\
    &=-2(m^2)^{\frac{D-4}{2}}\Gamma(2-\frac{D}{2})\neq0
\end{aligned}
\end{equation}

The contradiction indicates that the conjecture is not true in this case. We will find the ``bug'' for this example in the next sections.

\subsection{The cuts with Feynman prescription parameters kept}
The contradiction in Section \ref{sec:simple example} can be addressed more clearly if we keep the Feynman prescription parameters. Specifically, for an integral family with $n$ propagators $P_1,\cdots,P_n$, we define
\begin{equation}
    D_k= P_k - i \eta_k, \quad \Bar{D}_k= P_k +i \eta_k
\end{equation}
where $k=1,2,\cdots , n$ and $\eta_k$ will be taken as $0^+$ after all computation finished. We rewrite the Feynman integrals $I_{\alpha_1,\cdots,\alpha_n}$ in a finer way as
\begin{equation}
I_{\alpha_1,\cdots\alpha_n}=\lim_{\text{all }\eta_k\to0^+}\mathcal{I}_{\alpha_1,\cdots\alpha_n}
\end{equation}
where
\begin{equation}
    \mathcal{I}_{\alpha_1,\cdots\alpha_n}=\int \frac{\diff^{D}l_1}{i \pi^{\frac{D}{2}}}\cdots \frac{\diff^{D}l_L}{i \pi^{\frac{D}{2}}}\frac{1}{D_1^{\alpha_1}\cdots D_n^{\alpha_n}}
\end{equation}

We can define GUC through the ``mirroring'' the poles of the corresponding propagators. To do this, we define an $s$-mirrored integral, where $s$ is a set of indices, as

\begin{equation}
I_{\alpha_1,\cdots\alpha_n}|_{s\Mir}:=\lim_{\text{all }\eta_j\to0^+}\I_{\alpha_1,\cdots\alpha_n}|_{s\mir},
\end{equation}
where
\begin{equation}
f(\eta_1,\cdots,\eta_n)|_{s\mir}:=f(\eta_1,\cdots,\eta_n)|_{\eta_k \to -\eta_k,\, \text{for all } k\in s}
\end{equation}
for an expression $f(\eta_1,\cdots,\eta_n)$ containing the $\eta$'s. Notice that the mirror label with capital and lowercase of ``M'' have different definitions.

With the above, the GUC can be defined as

\begin{equation}
    I_{\alpha_1,\cdots\alpha_n}|_{\mathcal{C}\cut}=\lim_{\text{all }\eta_j\to0^+}\sum_{s \in \text{subsets of }\mathcal{C}}(-1)^{|s|}\I_{\alpha_1,\cdots\alpha_n}|_{s\mir},
\end{equation}
where $|s|$ is the number of elements in $s$. For example, if we cut one propagator, i.e. $s=\{k\}$, the above simplifies to
\begin{equation}
    I_{\alpha_1,\cdots\alpha_n}|_{\{k\}\cut}=\lim_{\text{all }\eta_j\to0^+}\int \frac{\diff^{D}l_1}{i \pi^{\frac{D}{2}}}\cdots \frac{\diff^{D}l_L}{i \pi^{\frac{D}{2}}}\Big(\frac{1}{D_1^{\alpha_1}\cdots D_n^{\alpha_n}}-\frac{1}{D_1^{\alpha_1}\cdots D_n^{\alpha_n}}\Big|_{\eta_k \to -\eta_k}\Big).
\end{equation}
This is consistent with what we are familiar with, if we admit the following relation\footnote{For some ill cases, one should be careful. For example, in cases that the function multiplied with $\delta$ function is singular at $P_k$=0, or it grows faster than $|P_k|$ when $|P_k|\to \infty$.}
\begin{equation}
    \lim_{\eta_k\to 0^+}(\frac{1}{D_k}-\frac{1}{\Bar{D}_k})=2 \pi i \delta(P_k)
\end{equation}

\subsection{The cause for the inconsistency}
With the Feynman prescription kept, the IBP relation in \eqref{eq: IBP example im bubble} is 
\begin{equation}\label{eq:full IBP of imbubble example}
    0=\int \frac{\diff^{D}l}{i \pi^{\frac{D}{2}}}\frac{\partial}{\partial l^{\mu}}\frac{p^\mu}{D_1D_2}
        =(\I_{0,2}-\I_{2,0}-i(\eta_2-\eta_1)(\I_{1,2}+\I_{2,1}))
\end{equation}
Applying $\{1\}\mir$ to \eqref{eq:full IBP of imbubble example}, we have
\begin{equation}\label{eq:full IBP with 1-mir of imbubble example}
    0=\int \frac{\diff^{D}l}{i \pi^{\frac{D}{2}}}\frac{\partial}{\partial l^{\mu}}\frac{p^\mu}{\bar{D}_1D_2}
        =(\I_{0,2}|_{\{1\}\mir}-\I_{2,0}|_{\{1\}\mir}-i(\eta_2+\eta_1)(\I_{1,2}|_{\{1\}\mir}+\I_{2,1}|_{\{1\}\mir}))
\end{equation}

In the above IBP relations, there are some terms that are proportional to Feynman prescription parameters, $\eta_k$'s, as
\begin{equation}
    -i(\eta_2+\eta_1)(\I_{1,2}|_{\{1\}\mir}+\I_{2,1}|_{\{1\}\mir}))
\end{equation}
These terms are the so-called \textit{hidden terms} in this paper. We will show that they are very closely related to the inconsistency problem. In the past, when we derive IBP relations, we always ignored hidden terms, since they are multiplied by $\eta_k$'s that will be taken to $\zp$. However, this is based on the assumption that the integrals, which multiplies $\eta_k$'s, are finite. We will show that this assumption is \textbf{not} always true after we apply ``mirror'' operations. For this example, we will show that
\begin{equation}
    \I_{1,2}|_{\{1\}\mir}+\I_{2,1}|_{\{1\}\mir}\sim \frac{1}{\eta_1+\eta_2}
\end{equation}
so that it cancels the ``vanishing'' coefficient $\eta_1+\eta_2$ in its front, making the whole term non-vanishing. 
\subsubsection{Hidden terms without mirroring $\eta_1$}
We can first look at the hidden terms in \eqref{eq:full IBP of imbubble example}. Without losing generality, we define
\begin{equation}\label{eq:l specify}
    l=(l^0,l^1,\vec{z}^{\,(D-2)})
\end{equation}
\begin{equation}\label{eq:p specify}
    p=(|\vec{p}\,|,|\vec{p}\,|,\vec{0}^{(D-2)})
\end{equation}
Integrating over $l^0$, we have
\begin{equation}
    \I_{1,2}+\I_{2,1}=\int\frac{\diff^{D-1}\vl}{i \pi^{\frac{D}{2}}}\B_{12},
\end{equation}
where
\begin{equation}\label{eq:B12}
   \B_{12}= \frac{i \pi  r_{12}}{2 r_1^2 r_2^2 (|\vp\,|^2-r_{12}^2)} \Big(1+\frac{2 |\vp\,|^2}{|\vp\,|^2-r_{12}^2}-\frac{r_{12}^2}{r_1 r_2}\Big)
\end{equation}
and 
\begin{equation}
    r_1=\sqrt{|\vl\,|^2+m^2+i \eta_1},\quad r_2=\sqrt{|\vl+\vp\,|^2+m^2+i \eta_2},\quad r_{12}=r_1+r_2.
\end{equation}
We expect that the integration of $B_{12}$ over $l^1$ is finite when $\eta_1$ and $\eta_2$ vanishes, because we have
\begin{equation}\label{eq:B12 limit}
    \lim_{l^1\to \pm\infty}\B_{12}|_{\eta_1=0,\eta_2=0} \sim \pm\frac{3i\pi }{4(l^1)^5}\to 0,
\end{equation}
and
\begin{equation}
    r_{12}>|\vp\,|, \quad \text{when }\eta_1,\eta_2=0,\quad\forall l^1\in(-\infty,+\infty).
\end{equation}

We can directly compute the integration over $l^1$ as\footnote{See Section \ref{app sec: derivation IntB12 definite} for its derivation.}
\begin{equation}\label{eq:IntB12 definite}
    \I_{1,2}+\I_{2,1}=\int\frac{\diff^{D-2}\vec{z}}{i \pi^{\frac{D}{2}}}\frac{i \pi }{ ( |\vec{z}\,|^2+m^2+i\eta _1)( |\vec{z}\,|^2+m^2+i\eta _2)}
\end{equation}
This is indeed finite. Thus, the term $(\eta_2-\eta_1)(\I_{1,2}+\I_{2,1})$ vanishes after taking $\eta_1,\eta_2=0^+$.
\subsubsection{Hidden terms with mirroring $\eta_1$}
For the hidden terms in \eqref{eq:full IBP with 1-mir of imbubble example}, the story is very different compared to the previous case. Again, using the gauge in \eqref{eq:l specify} and \eqref{eq:p specify}, and integrating over $l^0$ we have

\begin{equation}
    \I_{1,2}|_{\{1\}\mir}+\I_{2,1}|_{\{1\}\mir} = \int\frac{\diff^{D-1}\vl}{i \pi^{\frac{D}{2}}}\B_{\udon2}
\end{equation}
where
\begin{equation}\label{eq:Bu12}
   \B_{\udon2}= \frac{i \pi  r_{\udon2}}{2 r_\udon^2 r_2^2 (|\vp\,|^2-r_{\udon2}^2)} \Big(1+\frac{2 |\vp\,|^2}{|\vp\,|^2-r_{\udon2}^2}-\frac{r_{\udon2}^2}{r_\udon r_2}\Big)
\end{equation}
and
\begin{equation}
    r_\udon=-\sqrt{|\vl\,|^2+m^2-i \eta_1},\quad r_2=\sqrt{|\vl+\vp\,|^2+m^2+i \eta_2},\quad r_{\udon2}=r_\udon+r_2
\end{equation}
This time, the integration of $B_{\udon2}$ over $l^1$ is no longer convergent when $\eta_1,\eta_2\to \zp$, because
\begin{equation}\label{eq: Bu12 limit}
    \lim_{l^1\to \pm\infty}\B_{\udon2}|_{\eta_1=0,\eta_2=0} = \pm\frac{i\pi }{|\vp\,|(|\vz\,|^2+m^2)^2}\neq0.
\end{equation}
Consequently, we should keep $\eta_1$ and $\eta_2$ as regulators of the divergences while computing the integral. Otherwise, the expression becomes ill. 

We now compute the integration of $\B_\uot$ over $l^1$ with $\eta$'s kept. The result is\footnote{See Section \ref{app sec: derivation IntBu12 definite} for its derivation.}

\begin{equation}\label{eq:IntBu12 definite}
    \I_{1,2}|_{\{1\}\mir}+\I_{2,1}|_{\{1\}\mir} = \int\frac{\diff^{D-2}\vec{z}}{i \pi^{\frac{D}{2}}}\frac{- \pi  (2  (| \vec{z}\, | ^2+m^2)-i\eta _1+i\eta _2)}{(\eta _1+\eta _2) (| \vec{z}\, | ^2+m^2-i \eta _1) (| \vec{z}\, | ^2+m^2+i
   \eta _2)}
\end{equation}

The above results is indeed divergent with a pole proportional to $\frac{1}{\eta_1+\eta_2}$. This exactly cancels the ``vanishing'' coefficient $\eta_1+\eta_2$ in its front. Thus, the term is finite:
\begin{equation}\label{eq:conpemsate term}
    \lim_{\eta_1,\eta_2\to0^+}i({\eta_1+\eta_2})(\I_{1,2}|_{\{1\}\mir}+\I_{2,1}|_{\{1\}\mir}) = \int\frac{\diff^{D-2}\vec{z}}{i \pi^{\frac{D}{2}}}\frac{-2\pi i }{  (| \vec{z}\, | ^2+m^2)}=-2(m^2)^{\frac{D-4}{2}}\Gamma(2-\frac{D}{2})
\end{equation}
In the above, we took $\eta_1$ and $\eta_2$ as $0^+$ before the integration over $\vz$, since the integral is convergent after the pole $\frac{1}{\eta_1+\eta_2}$ is canceled. With the result in \eqref{eq:conpemsate term}, we can subtract \eqref{eq:full IBP of imbubble example} and \eqref{eq:full IBP with 1-mir of imbubble example} while taking $\eta_1, \eta_2=0^+$, to obtain
\begin{equation}\label{eq:imbubble example final correct IBP 1-cut}
    0=I_{0,2}|_{\Cut{1}}-I_{2,0}|_{\Cut{1}}-2(m^2)^{\frac{D-4}{2}}\Gamma(2-\frac{D}{2})
\end{equation}
Since $I_{0,2}|_{\Cut{1}}=0$, \eqref{eq:imbubble example final correct IBP 1-cut} agrees very well with \eqref{eq:1-cut spec calculation}. Thus, we can conclude that the hidden terms that we have previously ignored is the cause of the ``inconsistency'' of IBP relations, at least in this example.

\subsection{Comments}\label{sec:comments in sec 2}
The inconsistency in this example comes from the fact that the two propagators $P_1$ and $P_2$ are identical in the sense of integration over loop momenta, although they are different apparently. This kind of identical relation is not the symmetry relation such that $I_{\alpha_1,\alpha_2}=I_{\alpha_2,\alpha_1}$. It instead leads to an even stronger relation such that $I_{\alpha_1,\alpha_2}=I_{\alpha_1+\alpha_2,0}$. This can be interpreted by the fact that the integral family depends only on $m^2$, setting $p\mapsto \lambda p$ does not change the integral. When we set $\lambda=0$, $P_2$ comes to $P_1$. This leads to divergence while taking $\Cut{1}$ since, for example
\begin{equation}
    \lim_{\eta_1\to 0^+}(\I_{\alpha_1,\alpha_2}|_{\Cut{1}})\propto\int \diff^{D}l \frac{\delta^{(\alpha_1-1)}(P_1)}{(P_1-i \eta_2)^{\alpha_2}} = \int \diff^{D}l \frac{\delta^{(\alpha_1-1)}(P_1)}{(P_1-i \eta_2)^{\alpha_2}}\propto\frac{1}{\eta_2^{\alpha_1+\alpha_2-1}}
\end{equation}

This is consistent with what we observed in \eqref{eq:IntBu12 definite} after taking $\eta_1\to0^+$. Notice that, the singularity comes from the contribution where the poles $-i \eta_1$ and $i \eta_2$ located on the different side of the real axis, introducing a pinch singularity taking the limit $\eta\to0$.



In general, for a sector $S$, if the integral $I_{\alpha_1,\cdots,\alpha_n}$ on this sector is unchanged after we do the following replacement
\begin{equation}\label{eq: hidden relation}
    P_i= f(P_1,\cdots,P_{i-1},P_{i+1},\cdots,P_n)
\end{equation}
in the integrand, where $f$ satisfies $f(0,0,\cdots,0)=0$, cutting the propagators other than $P_i$ will introduce a divergence related to the Feynman prescriptions. These divergence could cancel with the coefficients in their front which vanishes when the Feynman prescriptions are set to $\zp$. In this paper, we call the relations like \eqref{eq: hidden relation} as \textit{hidden relations}.

\section{Searching for linear hidden relations}\label{sec: search linear hr}
As commented in Section \ref{sec:comments in sec 2}, the existence of hidden relations may lead to hidden terms in IBP relations, making the IBP breaks (if the hidden terms are erased) under the cuts. Thus, finding these relations is helpful to detect possible inconsistencies when using the spanning cuts method. In this section, we demonstrate an algorithm to find linear hidden relations using Feynman parameterization.

To derive Feynman parameterization, one will introduce a linear combination of Feynman parameters and propagators as 
\begin{equation}\label{eq: FP denomenator}
    \sum_{i=1}^n x_i P_i.
\end{equation}
If a linear hidden relation exists as
\begin{equation}
    P_i= \sum_{j=1,j\neq i}^n c_j P_j,
\end{equation}
\eqref{eq: FP denomenator} becomes
\begin{equation}
    \sum_{j=1,j\neq i}^n (x_j+c_j x_i) P_j.
\end{equation}
Consequently, the Symanzik polynomials (or $G=U+F$, the Lee-Pomeransky polynomial \cite{Lee:2013hzt,Lee:2013mka,Lee:2014tja}) should rely on $n-1$ variables, i.e. $x_j+c_j x_i$ for $j\neq i$. Then, we will have
\begin{equation}\label{eq: LP pol tangent relation}
    \Big(-\frac{\partial}{\partial x_i}+\sum_{j=1,j\neq i}^n c_j \frac{\partial}{\partial x_j}\Big)G=0
\end{equation}
Equivalently, we can find the linear hidden relations by search for annihilating differential vectors shown in \eqref{eq: LP pol tangent relation}. The algorithm is shown as Algorithm \ref{algorithm: finding linear hidden relations}.

\begin{algorithm}[htpb]
  \SetAlgoLined
  \KwIn{Lee-Pomeransky polynomial $G$, sector $S$}
  $G_s=G|_{\{x_i\to0|i\notin S\}}$.\\
  List dG, dG$_i$=$\frac{\partial G_s}{\partial x_i}$, for $i\in S$.\\
  List $t_j$=all terms in list dG. \\
  Matrix $M$, $M_{ij}=$ coefficient of $t_j$ in dG$_i$\\
  Matrix $N$=NullSpace(Transpose($M$))\\
  \KwOut{$N$ (or corresponding linear combinations of $x_i$)}
  \caption{The annihilating differential vectors in a sector $S$}
  \label{algorithm: finding linear hidden relations}
\end{algorithm}

We provide an implementation of this algorithm in \texttt{HiddenRelations.wl} in the following GitHub repo.

\begin{center}
    \url{https://github.com/Wu-Zihao/IBP-spc-consistency-problem-studies}
\end{center}

As an example, we applied Algorithm \eqref{algorithm: finding linear hidden relations} to the following integral family. Its propagators are
\begin{equation}
\begin{aligned}
    &P_1=l_1^2-m^2,\quad
    P_2=(l_1+p_1)^2-m^2,\quad
    P_3=(l_1+p_1+p_2)^2-m^2,\\
    &P_4=(l_2-p_1-p_2)^2-m^2,\quad
    P_5=(l_2+p_4)^2-m^2,\quad
    P_6=(l_2)^2-m^2,\\
    &P_7=(l_1+l_2)^2-m^2,\quad
    P_8=(l_1+p_4)^2,\quad
    P_9=(l_2+p_1)^2.
\end{aligned}
\end{equation}
The kinematic conditions are 
\begin{equation}
\begin{aligned}
    &p_1^2=p_2^2=p_4^2=0,\quad
    p_1\cdot p_2=\frac{s}{2},\quad
    p_1\cdot p_4=\frac{t}{2},\quad
    p_2\cdot p_4=\frac{-s-t}{2}.
\end{aligned}
\end{equation}

\begin{figure}[hbtp]
\centering
\subfloat[$\{1,2,4\}$]{

\begin{tikzpicture}[scale=0.9]
\draw[line width=2pt] (0,0) arc (120:60:2);
\draw[line width=2pt] (0,0) arc (-120:-60:2);
\draw[line width=2pt] (2,0) arc (135:180:0.707);
\draw[line width=2pt] (2,0) arc (45:0:0.707);
\draw[line width=2pt] (2-0.207,-0.49) arc (-180:0:0.207);

\draw(3,0.5)--(2,0);
\draw(3,0)--(2,0);
\draw(-1,0)--(0,0);
\draw(2,0)--(3,-0.5);
\node[font=\large\bfseries] at (-1.5,0){$1$};
\node[font=\large\bfseries] at (3.5,0){$3$};
\node[font=\large\bfseries] at (3.5,1){$2$};
\node[font=\large\bfseries] at (3.5,-1){$4$};
\node[color=blue] at (1,0.6){$P_1$};
\node[color=blue] at (1,-0.65){$P_2$};
\node[color=blue] at (2,-1.05){$P_4$};

\end{tikzpicture}
\label{subfig:imdb124}
}
\hspace{5pt}
\subfloat[$\{2,3,4,5\}$]{

\begin{tikzpicture}[scale=0.9]
\draw[line width=2pt] (0,0) arc (120:60:2);
\draw[line width=2pt] (0,0) arc (-120:-60:2);
\draw[line width=2pt] (2,0) arc (120:60:2);
\draw[line width=2pt] (2,0) arc (-120:-60:2);
\draw(2,1)--(2,-1);

\draw(-1,0)--(0,0);
\draw(5,0)--(4,0);
\node[font=\large\bfseries] at (-1.5,0){$3$};
\node[font=\large\bfseries] at (5.5,0){$2$};
\node[font=\large\bfseries] at (2,1.5){$1$};
\node[font=\large\bfseries] at (2,-1.5){$4$};

\node[color=blue] at (1,0.6){$P_4$};
\node[color=blue] at (1,-0.65){$P_5$};
\node[color=blue] at (3,0.6){$P_2$};
\node[color=blue] at (3,-0.65){$P_3$};
\end{tikzpicture}
\label{subfig:imdb2345}
}\\
\hspace{5pt}
\subfloat[$\{1,2,6,7\}$]{

\begin{tikzpicture}[scale=0.9]

\draw[line width=2pt] (0,0) arc (150-90:0:2);
\draw[line width=2pt] (0,0) arc (-90-90:-30-90:2);
\draw[line width=2pt](-1,-1.732)--(0,0);
\draw[line width=2pt](-1,-1.732)--(1,-1.732);
\draw(-1,-1.732)--(-1-1.732*0.5,-1.732-0.5);
\draw(1,-1.732)--(1+1.732*0.5,-1.732-0.5);
\draw(1,-1.732)--(1+0.5,-1.732-0.5*1.732);
\draw(1,-1.732)--(1+1,-1.732);

\node[font=\large\bfseries] at (-1-1.732*0.75,-1.732-0.75){$1$};
\node[font=\large\bfseries] at (1+1.732*0.75,-1.732-0.75){$3$};
\node[font=\large\bfseries] at (1+1.5,-1.732){$2$};
\node[font=\large\bfseries] at (1+0.75,-1.732-0.75*1.732){$4$};

\node[color=blue] at (-1,-0.7){$P_1$};
\node[color=blue] at (1,-0.65){$P_7$};
\node[color=blue] at (0,-1.25){$P_6$};
\node[color=blue] at (0,-2.25){$P_2$};

\end{tikzpicture}
\label{subfig:imdb1267}
}
\hspace{5pt}
\subfloat[$\{1,6,8,9\}$]{

\begin{tikzpicture}[scale=0.9]
\draw[line width=2pt] (0,0) arc (120:60:2);
\draw (0,0) arc (-120:-60:2);
\draw[line width=2pt] (2,0) arc (120:60:2);
\draw (2,0) arc (-120:-60:2);
\draw(2,1)--(2,-1);

\draw(-1,0)--(0,0);
\draw(5,0)--(4,0);
\node[font=\large\bfseries] at (-1.5,0){$1$};
\node[font=\large\bfseries] at (5.5,0){$4$};
\node[font=\large\bfseries] at (2,1.5){$2$};
\node[font=\large\bfseries] at (2,-1.5){$3$};

\node[color=blue] at (1,0.6){$P_6$};
\node[color=blue] at (1,-0.65){$P_9$};
\node[color=blue] at (3,0.6){$P_1$};
\node[color=blue] at (3,-0.65){$P_8$};
\end{tikzpicture}
\label{subfig:imdb1689}
}

\caption{Diagrams of the representative sectors with linear hidden relation(s). }
\label{fig:imdb sub sectors}
\end{figure}
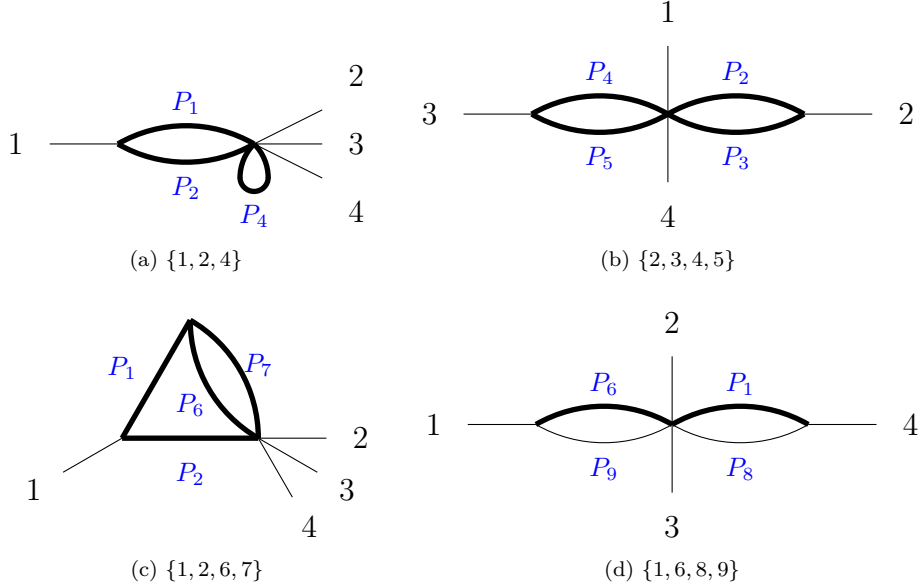

After searching all non-zero sectors, we found in total 65 sectors with annihilating differential vectors shown in \eqref{eq: LP pol tangent relation}. We demonstrate some representative ones among them in Table \ref{tab: massive dbox sectors with hidden relations}. The corresponding diagrams are shown in Figure \ref{fig:imdb sub sectors}.
\begin{table}[h]
    \centering
    \begin{tabular}{|c|c|c|c|}
    \hline
        sector & ann. diff. vector(s) & hidden relation(s) & apparent value(s)\\
    \hline
    \{1,2,4\}&$\partial_1 - \partial_2$ & $P_1-P_2$ & $ -2l_1\cdot p_1$\\
    \hline
    \{2,3,4,5\}&$\partial_2 - \partial_3, \,\partial_5 - \partial_4 $ & $P_2-P_3,\, P_4-P_5$& $ -2(l_1+p_1)\cdot p_2,\, 2 (l_2+p_4)\cdot p_3  $\\
    \hline
    \{1,2,6,7\}&$\partial_1 - \partial_2$ & $P_1-P_2$& $ -2l_1\cdot p_1 $\\
    \hline
    \{1,6,8,9\}&$\partial_1 - \partial_6-\partial_8 + \partial_9$ & $P_1-P_6-P_8+P_9$& $ 2(l_2\cdot p_1-l_1\cdot p_4) $\\
    \hline
    \end{tabular}
    \caption{Some representative sectors with linear hidden relation(s).}
    \label{tab: massive dbox sectors with hidden relations}
\end{table}

The cause of the inconsistencies is understandable for the sub diagram in Figure \ref{subfig:imdb124} because it is factorized with a factor of massive bubble shown in Figure \ref{fig:imbubble}. The sub diagram in Figure \ref{subfig:imdb2345} is similar. It has two factorized massive bubbles. So, it has two hidden relations. The sub diagram shown in Figure \ref{subfig:imdb1267} does not have such massive bubble factor, but one can also set $p_1\to 0$ since the diagram has only one massless external line. After doing so, $P_1$ and $P_2$ become identical, introducing divergence when cutting one of them. The sub diagram shown in Figure \ref{subfig:imdb1689} is relatively more special. Its hidden relation consists of four terms rather than two. This means that it will show divergence if we cut three of the four propagators.  

We have noticed that, while the preparation of this paper, a very interesting paper \cite{Crisanti:2026rbc} was posted, studying the implications between spanning cuts inconsistency, magic relations, and the dimensions of critical varieties \cite{Lee:2013hzt} from the Lee-Pomeransky polynomials. In this work, the authors include codes to find sectors where the critical varieties are with high dimensions, indicating that this implies the existence of magic relations, which usually breaks under the cuts. Although the algorithm in \cite{Crisanti:2026rbc} and the Algorithm \ref{algorithm: finding linear hidden relations} in this paper are designed individually\footnote{We hope this is convincing, because the submission dates of the two papers are very close.} form different starting points, they appear to capture the same underlying mechanism. To specify this comment, we point out that in Algorithm \ref{algorithm: finding linear hidden relations}, if a linear hidden relation is found, we know that the number of independent variables in the Lee-Pomenransky polynomials will be decreased. This leads to a higher-dimensioned  critical variety. Thus, at least at linear level, this paper explains why the consistency problem is related to the dimension of critical variety. On the other hand, the studies in \cite{Crisanti:2026rbc} gives us a very strong hint on how to find non-linear hidden relations.

\section{Inconsistencies of external-massive diagrams}\label{sec: 2-loop special example}

The examples in previous sections show that the diagrams with inconsistencies have a common feature. That is, they have the only one massless external line, or is factorized and has such diagram(s) as factor(s) (we call this feature \textit{1EML} for convenience in this paper). However, this does not mean that other diagrams are safe. In this section, we give an example with massive external lines, which still has inconsistency. 

\begin{figure}[hbtp]
\centering
\begin{tikzpicture}[scale=1.2]

\draw[line width=2.5pt](-1,0) -- (0,0);
\draw[line width=2pt,color=dark gray](2,0) -- (0,0);
\draw[line width=2.5pt](3,0) -- (1.95,0);

\draw (1,1.732)--(0,0);
\draw[line width=2.5pt] (1,1.732) arc (-180:-180+60:2);
\draw[line width=2.5pt] (1,1.732) arc (60:0:2);

\node[font=\large\bfseries,color=dark gray] at (1,-0.3){$m^2$};
\node[font=\large\bfseries] at (-1.3,0){$p$};
\node[font=\large\bfseries] at (3.3,0){$p$};
\node[font=\large\bfseries] at (-0.5,-0.25){$s$};
\node[font=\large\bfseries] at (2.5,-0.25){$s$};
\node[font=\large\bfseries] at (1.1,0.65){$s$};
\node[font=\large\bfseries] at (1.94,1){$s$};

\end{tikzpicture}
\caption{A two-loop two-point example with external massive lines}
\label{fig:two-loop special example}
\end{figure}
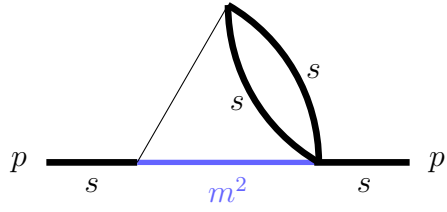

The diagram of the example is shown in Figure \ref{fig:two-loop special example}. The propagators of the example are
\begin{equation}
    P_1=l_1^2,\quad
    P_2=(l_1+p)^2-m^2,\quad
    P_3=l_2^2-s,\quad
    P_4=(l_1+l_2)^2-s,\quad
    P_5=(l_2+p)^2.
\end{equation}
$P_5$ is an irreducible scalar product. The kinematic condition is $p^2=s$, where $s\neq m^2$.

For this example, we can write down an IBP relation as 
\begin{equation}\label{eq:2L IBP with h.t.}
\begin{aligned}
    0=&\int \frac{\diff^{D}l_1}{i \pi^{\frac{D}{2}}} \frac{\diff^{D}l_2}{i \pi^{\frac{D}{2}}}\frac{\partial}{\partial l_2^\mu}\frac{l_1^\mu}{D_1 D_2 D_3 D_4}\\
    =&\int \frac{\diff^{D}l_1}{i \pi^{\frac{D}{2}}} \frac{\diff^{D}l_2}{i \pi^{\frac{D}{2}}}\frac{(D_3-D_4)(D_3+D_4-D_1)+i(\eta_3-\eta_4-\eta_1)D_3+i(\eta_3-\eta_4+\eta_1)D_4}{D_1 D_2 D_3^2 D_4^2}\\
    =&(\I_{1,1,0,2}-\I_{1,1,2,0}-\I_{0,1,1,2}+\I_{0,1,2,1}+i(\eta_3-\eta_4-\eta_1)\I_{1,1,1,2}+i(\eta_3-\eta_4+\eta_1)\I_{1,1,2,1})
\end{aligned}
\end{equation}
with hidden terms
\begin{equation}\label{eq:H34}
 \HT_{34}=i(\eta_3-\eta_4)(\I_{1,1,1,2}+\I_{1,1,2,1})-i\eta_1(\I_{1,1,1,2}-\I_{1,1,2,1})
\end{equation}
If we apply $\Cut{1,3}$ on \eqref{eq:2L IBP with h.t.} and blindly erase the hidden terms proportional to $\eta$'s, we will get
\begin{equation}
    I_{1,1,2,0}|_{\Cut{1,3}}=0
\end{equation}
However this is wrong. We can compute it as\footnote{See Appendix \ref{app sec: 2L} for its derivation.}
\begin{equation}\label{eq:I1120 1,3-cut direct integration}
\begin{aligned}
    I_{1,1,2,0}|_{\Cut{1,3}}=&-\lim_{\eta_2\to\zp}(2\pi i)^2\int \frac{\diff^D l_1\diff^D l_2}{(i\pi^{\frac{D}{2}})^2}\frac{\delta(P_1)\delta^\prime(P_3)}{D_2}\\
    =&4\pi e^{i \phi}\cot{(\frac{\pi D}{2})}\Gamma(3-D)\frac{ (s-m^2)^{D-3}}{s}\neq 0,
\end{aligned}
\end{equation}
where $\phi$ is a real parameter depending on the sign of $(s-m^2)$. This result shows that the inconsistency appears in this example.

\begin{remark}\label{remark 1}
    One might wonder whether we can find a hidden relation to explain the inconsistency in this example. After applying Algorithm \ref{algorithm: finding linear hidden relations}, we fond that there does not exist linear hidden relations for this sector.
\end{remark}

Although we did not find the linear hidden relations, we still have an intuitive understanding for the cause of inconsistency of this example. Indeed, the diagram itself is not with feature \textit{1EML}. However, after cutting $P_1$ (the massless propagator), making it on-shell, the sub-diagram (the massive bubble formed by $P_3$ and $P_4$) is with the feature \textit{1EML}. We expect that under $\{1\}-$cut, $P_3$ and $P_4$ should be related by some relations, similar to the linear hidden relations defined in this paper.

There are still many works that can be done for this example. Firstly, one may want to reveal explicitly the relations between $P_3$ and $P_4$. Secondly, one may want to compute the hidden terms in this example to see whether it coincides \eqref{eq:I1120 1,3-cut direct integration}. We are leaving these works for future studies.

\section{Summary}\label{sec: conclusion}

In this paper, we have identified one of the explanations of the spanning cuts consistency problem.  We claim that, the conjecture \textit{IBP relations is unchanged under the cuts}, is some times incorrect, if one blindly erases the hidden terms proportional to the ``vanishing'' Feynman prescription parameters.

We provided several examples to demonstrate this mechanism. In some of them,
the inconsistencies can be explained by linear hidden relations (like those in Section \ref{sec:simple example} and Section \ref{sec: search linear hr}), while the others cannot (like the one in Section \ref{sec: 2-loop special example}). We have also observed that the inconsistencies usually appear together with the \textit{1EML} feature. This feature, in many cases, indicates the presence of hidden relations.

In Section \ref{sec: search linear hr}, we provided an algorithm together with an implementation to find linear hidden relations. We have commented in Section \ref{sec: search linear hr} that, inspired by the very recent work \cite{Crisanti:2026rbc} on this topic, the existence of hidden relations may be closely related to the dimensions of the critical varieties from the Lee-Pomeransky polynomial.

As a disclaimer, although this paper has demonstrated plenty of features that lead to inconsistency, we \textbf{do not} claim that these features exhaust all possible reasons of this problem. For example, we did not prove that diagrams without \textit{1EML} feature are safe. Nor have we proved that the families free of hidden relations are safe. Thus, there are still plenty of studies required on this topic. We leave these questions for future work.

\section*{Acknowledgement} 
We thank Bo Feng, Yuanche Liu, Yan-Qing Ma, Johann Usovitch, Ellis Ye Yuan and Yang Zhang for important discussions. 
Zihao Wu is supported by South China Normal University through grant No. 43102673.
Yingxuan Xu is supported by the Deutsche
Forschungsgemeinschaft (DFG, German Research Foundation) under grant 396021762 - TRR 257.

\appendix

\section{Derivations}

In this section, we collect the derivation details of some of the equations in this paper.

\subsection{Equation \eqref{eq:IntB12 definite}}\label{app sec: derivation IntB12 definite}

The derivation of equation \eqref{eq:IntB12 definite} is to compute the integration of $\B_{12}$ defined in \eqref{eq:B12}. To make it convenient for the readers, we copy it here
\begin{equation}\label{eq:B12 appendix}
   \B_{12}= \frac{i \pi  r_{12}}{2 r_1^2 r_2^2 (|\vp\,|^2-r_{12}^2)} \Big(1+\frac{2 |\vp\,|^2}{|\vp\,|^2-r_{12}^2}-\frac{r_{12}^2}{r_1 r_2}\Big)
\end{equation}
The indefinite integration of $\B_{12}$ over $l^1$ can be obtained, as\footnote{In this paper, we add a subscript ``id'' in the integration label, as $\idint$, in order to emphasize that this is an indefinite integration.} 
\begin{equation}\label{eq:IntB12 indefinite}
    \beta_{12}:=\idint \B_{12} \diff l^1=-\frac{i \pi |\vp\,|}{2 Q_{12}}\Big(
    \frac{3q_1+4(l^1)^2-F_{12}}{r_1 q_1}- \frac{3q_2+4(|\vp\,|+l^1)^2-F_{12}}{r_2 q_2}
    \Big)
\end{equation}
where

\begin{equation}\label{eq:rep for beta12}
\begin{aligned}
   & q_1=|\vz\,|^2+m^2+i \eta_1, \\&
     q_2=|\vz\,|^2+m^2+i \eta_2, \\&
     Q_{12}=(\eta_1-\eta_2)^2+4 F_{12} |\vp\,|^2,\\&
     F_{12}=q_1+i(\eta_1-\eta_2)\frac{ l^1}{|\vp\,|}.
\end{aligned}
\end{equation}

In order to get the definite integration over $l^1$, we need to analyze the poles and branch cuts of the integrand $\B_{12}$. In the denominator of the integrand, in apparent,  $|\vp\,|^2-r_{12}^2$ contributes pole(s), and $r_1$ and $r_2$ contribute branch points and branch cuts. 

\paragraph{Pole(s) }
We analyze the pole(s). Let 
\begin{equation}
    |\vp\,|^2-r_{12}^2=0
\end{equation}
The L.H.S of the above equation contains a term with square roots $(-2 r_1 r_2)$. To remove it, consider
\begin{equation}\label{eq:A5 Q12}
 (|\vp\,|^2-r_{12}^2)( (|\vp\,|^2-(r_1-r_2)^2))=-Q_{12}.
\end{equation}
Now, $Q_{12}$ is linear in $l^1$. For the equation $Q_{12}=0$, if $\eta_1=\eta_2$, there is no solution. Otherwise, there is a solution
\begin{equation}\label{eq:solution Q12=0}
    l^1=\frac{i}{4}\Big(\frac{\eta_1-\eta_2}{|\vp\,|}+\frac{4|\vp\,|(|\vz\,|^2+m^2+ i \eta_1)}{\eta_1-\eta_2}\Big)
\end{equation}
This solution, of course, can make only one bracket vanish in the L.H.S. of \eqref{eq:A5 Q12}. To figure out which is the one, we substitute the solution \eqref{eq:solution Q12=0} into $r_1$ and $r_2$, we have
\begin{equation}\label{eq:r1 substituted by solution of Q12=0}
    r_1=\frac{1}{4}\sqrt{-\frac{\Big((\eta_1-\eta_2)^2-4\vpnorm^2(\vznorm^2+m^2+i \eta_1)\Big)^2}{(\eta_1-\eta_2)^2\vpnorm^2}}
\end{equation}
and
\begin{equation}\label{eq:r2 substituted by solution of Q12=0}
    r_2=\frac{1}{4}\sqrt{-\frac{\Big((\eta_1-\eta_2)^2-4\vpnorm^2(\vznorm^2+m^2+i \eta_2)\Big)^2}{(\eta_1-\eta_2)^2\vpnorm^2}}.
\end{equation}
Since, by convention, the square root is with a branch cut along the negative real axis, $r_1$ and $r_2$ must have positive real parts. Thus, we have
\begin{equation}
    r_1={-\frac{(\eta_1-\eta_2)^2-4\vpnorm^2(\vznorm^2+m^2+i \eta_1)}{4i|\eta_1-\eta_2|\vpnorm}}
\end{equation}
and
\begin{equation}
    r_2={-\frac{(\eta_1-\eta_2)^2-4\vpnorm^2(\vznorm^2+m^2+i \eta_2)}{4i|\eta_1-\eta_2|\vpnorm}}.
\end{equation}
Obviously, 
\begin{equation}
    r_1-r_2=\vpnorm\text{Sign}(\eta_1-\eta_2)
\end{equation}
Thus, the solution \eqref{eq:solution Q12=0} makes
\begin{equation}\label{eq:p^2-(r1-r2)^2=0 when Q12=0}
    \vpnorm^2-(r_1-r_2)^2=0.
\end{equation}
Since $r_1r_2\neq 0$, this solution cannot make $\vpnorm^2-r_{12}^2=0$. 

From the above, we know that the denominator $\vpnorm^2-r_{12}^2=0$ does not contribute poles for $\B_{12}$, both when $\eta_1=\eta_2$ and $\eta_1\neq \eta_2$.

\paragraph{Branch cuts}

The branch cuts introduced by $r_i$ can be described by 
\begin{equation}
    r_i^2=-\R_i,
\end{equation}
where $i=1,2$ and $\R_i\geq 0$ are real parameters describing the branch cuts and branch points (when $\R_i=0$). Solving the above equations, we get the branch cuts from $r_1$ are
\begin{equation}\label{eq:branch cuts r1}
    l^1=\pm\frac{1}{\sqrt{2}}\Big(\K_1-\frac{i\eta_1}{\K_1}\Big),
\end{equation}
and the branch cuts from $r_2$ are
\begin{equation}\label{eq:branch cuts r2}
    l^1=-\vpnorm\pm\frac{1}{\sqrt{2}}\Big(\K_2-\frac{i\eta_2}{\K_2}\Big),
\end{equation}
where
\begin{equation}\label{eq:Ki definition}
    \K_i=\sqrt{\sqrt{(\vznorm^2+m^2+\R_i)^2+\eta_i^2}-(\vznorm^2+m^2+\R_i)}
\end{equation}

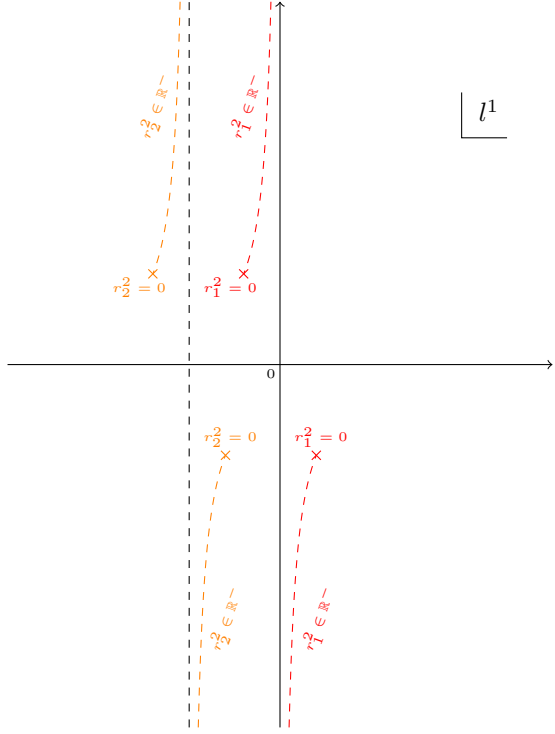
\begin{figure}[hbtp]
\centering
\begin{tikzpicture}[scale=1.2]

\draw[dashed,color=red,domain=0.1:0.4, variable=\t] plot ({\t}, {-0.4/\t});
\draw[dashed,color=red,domain=-0.1:-0.4, variable=\t] plot ({\t}, {-0.4/\t});

\draw[color=red](0.4,-1)--(0.45,-1.05);
\draw[color=red](0.4,-1)--(0.35,-0.95);
\draw[color=red](0.4,-1)--(0.45,-0.95);
\draw[color=red](0.4,-1)--(0.35,-1.05);
\draw[color=red](-0.4,1)--(-0.45,1.05);
\draw[color=red](-0.4,1)--(-0.35,0.95);
\draw[color=red](-0.4,1)--(-0.45,0.95);
\draw[color=red](-0.4,1)--(-0.35,1.05);

\draw [dashed] (-1,-4)--(-1,4);

\draw[dashed,color=orange,domain=0.1:0.4, variable=\t] plot ({-1+\t}, {-0.4/\t});
\draw[dashed,color=orange,domain=-0.1:-0.4, variable=\t] plot ({-1+\t}, {-0.4/\t});

\draw[color=orange](-0.6,-1)--(-0.55,-1.05);
\draw[color=orange](-0.6,-1)--(-0.65,-0.95);
\draw[color=orange](-0.6,-1)--(-0.55,-0.95);
\draw[color=orange](-0.6,-1)--(-0.65,-1.05);
\draw[color=orange](-1.4,1)--(-1.45,1.05);
\draw[color=orange](-1.4,1)--(-1.35,0.95);
\draw[color=orange](-1.4,1)--(-1.45,0.95);
\draw[color=orange](-1.4,1)--(-1.35,1.05);

\node[font=\tiny] at (-0.1,-0.1) {0};

\node[orange, font=\tiny] at (-0.55,-0.8) {$r_2^2 = 0$};
\node[orange, font=\tiny] at (-1.55, 0.85) {$r_2^2 = 0$};
\node[orange, font=\tiny,rotate=70] at (-0.58,-2.8) {$r_2^2\in\mathbb{R}^-$};
\node[orange, font=\tiny,rotate=70] at (-1.35, 2.85) {$r_2^2\in\mathbb{R}^-$};

\node[red, font=\tiny] at (0.45, -0.8) {$r_1^2 = 0$};
\node[red, font=\tiny] at (-0.55, 0.85) {$r_1^2 = 0$};
\node[red, font=\tiny, rotate=70] at (0.44, -2.8) {$r_1^2 \in\mathbb{R}^-$};
\node[red, font=\tiny, rotate=70] at (-0.35, 2.85) {$r_1^2 \in\mathbb{R}^-$};

\draw[->] (-3,0)--(3,0);
\draw[->] (0,-4)--(0,4);

\draw(2,3)--(2,2.5)--(2.5,2.5);

\node[ rotate=0] at (2.3, 2.8) {$l^1$};

\end{tikzpicture}
\caption{Branch points and branch cuts of $\B_{12}$}
\label{fig:Branch points and branch cuts of B12}
\end{figure}

According to the above expressions, none of the branch cuts crosses the real axis. They are shown in Figure \ref{fig:Branch points and branch cuts of B12}.

\paragraph{The integration }
From the analysis of pole(s) and branch cuts above, we can now safely take the limit of $\beta_{12}$ defined in \eqref{eq:IntB12 indefinite} at $l^1\to\pm\infty$, to get the definite integration of $\B_{12}$ along the real axis. At such limits, if $\eta_1\neq\eta_2$, we have
\begin{equation}
    r_1\to |l^1|,\quad r_2\to |l^1|,
\end{equation}
and
\begin{equation}
    F_{12}\to i(\eta_1-\eta_2)\frac{l^1}{\vpnorm},\quad 
    Q_{12}\to 4 i(\eta_1-\eta_2)\vpnorm{l^1}.
\end{equation}
Thus,
\begin{equation}\label{eq:beta12 limit}
\begin{aligned}
    \beta_{12}&\to-\frac{2  \pi i|\vp\,|(l^1)^2}{ Q_{12}}\Big(
    \frac{1}{r_1 q_1}- \frac{1}{r_2 q_2}
    \Big)\\&
    \to -\frac{  \pi l^1}{2 (\eta_1-\eta_2)|l^1|}\Big(
    \frac{1}{q_1}- \frac{1}{ q_2}
    \Big)
    =\frac{i\pi}{2 q_1 q_2}\Sign(l^1).
\end{aligned}
\end{equation}
If $\eta_1=\eta_2$, we have
\begin{equation}\label{eq: noname A.20}
    q_1=q_2=F_{12}, \quad Q_{12}=4 \vpnorm^2 q_1,
\end{equation}
We consider the factor in the bracket of $\beta_{12}$ in \eqref{eq:IntB12 indefinite}, as follows,
\begin{equation}\label{eq:beta12 factor in bracket when eta1=eta2}
\begin{aligned}
    &\frac{3q_1+4(l^1)^2-F_{12}}{r_1 q_1}- \frac{3q_2+4(|\vp\,|+l^1)^2-F_{12}}{r_2 q_2}\\&
    =\Big(2+\frac{4(l^1)^2}{q_1}\Big)\Big(\frac{1}{r_1}-\frac{1}{r_2}\Big) - \frac{8\vpnorm l^1+4\vpnorm^2}{r_2 q_1}\\&
    =\Big(2+\frac{4(l^1)^2}{q_1}\Big)\frac{2\vpnorm l^1 +\vpnorm^2}{r_1 r_2r_{12}} - \frac{8\vpnorm l^1+4\vpnorm^2}{r_2 q_1}\\&
    =\frac{2\vpnorm l^1 +\vpnorm^2}{q_1r_1 r_2r_{12}}\Big(2q_1+4(l^1)^2-4r_1r_{12}\Big).
\end{aligned}
\end{equation}
Since
\begin{equation}
    r_{12}:=r_1+r_2\to 2|l^1|,
\end{equation}
we have
\begin{equation}
    \frac{2\vpnorm l^1 +\vpnorm^2}{q_1r_1 r_2r_{12}}\Big(2q_1+4(l^1)^2-4r_1r_{12}\Big)\to-\frac{4(l^1)^3\vpnorm}{q_1|l^1|^3}=-\frac{4\vpnorm}{q_1}\Sign(l^1).
\end{equation}
Using \eqref{eq:IntB12 indefinite} and \eqref{eq: noname A.20}, we have
\begin{equation}\label{eq:beta12 limit when eta1=eta2}
    \beta_{12}\to\frac{i\pi }{2q_1^2}\Sign(l^1).
\end{equation}
This agrees with \eqref{eq:beta12 limit} when taking $\eta_1=\eta_2$. So, \eqref{eq:beta12 limit} applies also to the cases when $\eta_1=\eta_2$. Using it, we can finally get
\begin{equation}
    \int \B_{12}\diff l^l=\lim_{l^1\to \infty}\beta_{12}-\lim_{l^1\to -\infty}\beta_{12}=\frac{i\pi}{q_1q_2}.
\end{equation}
It is \eqref{eq:IntB12 definite}.

\subsection{Equation \eqref{eq:IntBu12 definite}}\label{app sec: derivation IntBu12 definite} 
We compute the integration of $\B_{\udon2}$ defined in \eqref{eq:Bu12}. To make it convenient for the readers, we copy it here
\begin{equation}\label{eq:Bu12 appendix}
   \B_{\udon2}= \frac{i \pi  r_{\udon2}}{2 r_\udon^2 r_2^2 (|\vp\,|^2-r_{\udon2}^2)} \Big(1+\frac{2 |\vp\,|^2}{|\vp\,|^2-r_{\udon2}^2}-\frac{r_{\udon2}^2}{r_\udon r_2}\Big)
\end{equation}
Its indefinite integration is

\begin{equation}\label{eq:IntBu12 indefinite}
    \beta_{\udon2}:=\idint \B_{\udon2} \diff l^1=-\frac{i \pi |\vp\,|}{2 Q_{\udon2}}\Big(
    \frac{3q_\udon+4(l^1)^2-F_{\udon2}}{r_\udon q_\udon}- \frac{3q_2+4(|\vp\,|+l^1)^2-F_{\udon2}}{r_2 q_2}
    \Big)
\end{equation}
where

\begin{equation}\label{eq:rep for betau12}
\begin{aligned}
   & q_\udon=|\vz\,|^2+m^2-i \eta_1, \\&
     q_2=|\vz\,|^2+m^2+i \eta_2, \\&
     Q_{\udon2}=(\eta_1+\eta_2)^2+4 F_{\udon2} |\vp\,|^2,\\&
     F_{\udon2}=q_\udon-i(\eta_1+\eta_2)\frac{ l^1}{|\vp\,|}.
\end{aligned}
\end{equation}

Still, we analyze the pole(s) and branch cuts of $\B_{\udon2}$.

\paragraph{Pole(s)}
Similarly, we have 
\begin{equation}\label{eq:A Qu12}
 (|\vp\,|^2-r_{\udon2}^2)( (|\vp\,|^2-(r_\udon-r_2)^2))=-Q_{\udon2},
\end{equation}
which is linear in $l^1$. Since $\eta_1+\eta_2>0$, the solution of $Q_{\udon2}=0$ is 
\begin{equation}\label{eq:solution Qu12=0}
    l^1=-\frac{i}{4}\Big(\frac{\eta_1+\eta_2}{|\vp\,|}+\frac{4|\vp\,|(|\vz\,|^2+m^2- i \eta_1)}{\eta_1+\eta_2}\Big)
\end{equation}
To figure out which bracket in the L.H.S. of \eqref{eq:A Qu12} vanishes, we substitute \eqref{eq:solution Qu12=0} to $r_\udon$ and $r_2$ to get
\begin{equation}\label{eq:ru1 substituted by solution of Qu12=0}
    r_\udon=-\frac{1}{4}\sqrt{-\frac{\Big((\eta_1+\eta_2)^2-4\vpnorm^2(\vznorm^2+m^2-i \eta_1)\Big)^2}{(\eta_1+\eta_2)^2\vpnorm^2}}
\end{equation}
and
\begin{equation}\label{eq:r2 substituted by solution of Qu12=0}
    r_2=\frac{1}{4}\sqrt{-\frac{\Big((\eta_1+\eta_2)^2-4\vpnorm^2(\vznorm^2+m^2+i \eta_2)\Big)^2}{(\eta_1+\eta_2)^2\vpnorm^2}}.
\end{equation}
Also, considering the conventional branch cuts of the square root function, $r_\udon$ must have a negative real part, and $r_2$ must have a positive one. Thus, we have
\begin{equation}
    r_1=-\frac{(\eta_1+\eta_2)^2-4\vpnorm^2(\vznorm^2+m^2-i \eta_1)}{4i(\eta_1+\eta_2)\vpnorm}
\end{equation}
and
\begin{equation}
    r_2=-{\frac{(\eta_1+\eta_2)^2-4\vpnorm^2(\vznorm^2+m^2+i \eta_2)}{4i(\eta_1+\eta_2)\vpnorm}}.
\end{equation}
Consequently, 
\begin{equation}
    r_1-r_2=-\vpnorm
\end{equation}
Thus, the solution \eqref{eq:solution Qu12=0} makes
\begin{equation}\label{eq:p^2-(r1-r2)^2=0 when Qu12=0}
    \vpnorm^2-(r_\udon-r_2)^2=0.
\end{equation}
Since $r_\udon r_2\neq 0$, this solution cannot make $\vpnorm^2-r_{\udon2}^2=0$. i.e. the denominator $\vpnorm^2-r_{\udon2}^2$ does not contribute any poles for $\B_{\udon2}$

\paragraph{Branch cuts}
We describe the branch cuts introduced by $r_i$ by 
\begin{equation}
    r_i^2=-\R_i,
\end{equation}
where $i=\udon,2$ and $\R_i\geq 0$ are real parameters. The branch cuts from $r_\udon$ are
\begin{equation}\label{eq:branch cuts ru1}
    l^1=\pm\frac{1}{\sqrt{2}}\Big(\K_\udon+\frac{i\eta_1}{\K_\udon}\Big),
\end{equation}
where
\begin{equation}\label{eq:Ku1 definition}
    \K_\udon=\sqrt{\sqrt{(\vznorm^2+m^2+\R_\udon)^2+\eta_\udon^2}-(\vznorm^2+m^2+\R_\udon)}.
\end{equation}

The branch cuts from $r_2$ are already shown in \eqref{eq:branch cuts r2}. In Figure \ref{fig:Branch points and branch cuts of Bu12} we draw the branch cuts of $\B_{\udon2}$. None of them crosses the real axis. Thus, it is also safe to use $\beta_{\udon2}$ to compute the definite integration of $\B_{\udon2}$.

\begin{figure}[hbtp]
\centering
\begin{tikzpicture}[scale=1.2]

\draw[dashed,color=red,domain=0.1:0.4, variable=\t] plot ({\t}, {0.4/\t});
\draw[dashed,color=red,domain=-0.1:-0.4, variable=\t] plot ({\t}, {0.4/\t});

\draw[color=red](0.4,1)--(0.45,1.05);
\draw[color=red](0.4,1)--(0.35,0.95);
\draw[color=red](0.4,1)--(0.45,0.95);
\draw[color=red](0.4,1)--(0.35,1.05);
\draw[color=red](-0.4,-1)--(-0.45,-1.05);
\draw[color=red](-0.4,-1)--(-0.35,-0.95);
\draw[color=red](-0.4,-1)--(-0.45,-0.95);
\draw[color=red](-0.4,-1)--(-0.35,-1.05);

\draw[dashed,color=orange,domain=0.1:0.4, variable=\t] plot ({-1.5+\t}, {-0.4/\t});
\draw[dashed,color=orange,domain=-0.1:-0.4, variable=\t] plot ({-1.5+\t}, {-0.4/\t});

\draw[color=orange](-1.1,-1)--(-1.05,-1.05);
\draw[color=orange](-1.1,-1)--(-1.15,-0.95);
\draw[color=orange](-1.1,-1)--(-1.05,-0.95);
\draw[color=orange](-1.1,-1)--(-1.15,-1.05);
\draw[color=orange](-1.9,1)--(-1.95,1.05);
\draw[color=orange](-1.9,1)--(-1.85,0.95);
\draw[color=orange](-1.9,1)--(-1.95,0.95);
\draw[color=orange](-1.9,1)--(-1.85,1.05);

\draw [dashed] (-1.5,-4)--(-1.5,4);

\node[red, font=\tiny] at (0.45, 0.8) {$r_\udon^2 = 0$};
\node[red, font=\tiny,rotate=-30] at (-0.45, -0.85) {$r_\udon^2 = 0$};
\node[red, font=\tiny, rotate=-70] at (0.44, 2.8) {$r_\udon^2\in\mathbb{R}^-$};
\node[red, font=\tiny, rotate=-70] at (-0.35, -2.85) {$r_\udon^2\in\mathbb{R}^-$};

\node[orange, font=\tiny,rotate=30] at (-1.15, -0.78) {$r_2^2 = 0$};
\node[orange, font=\tiny] at (-2.05, 0.85) {$r_2^2 = 0$};
\node[orange, font=\tiny, rotate=70] at (-1.08, -2.8) {$r_2^2\in\mathbb{R}^-$};
\node[orange, font=\tiny, rotate=70] at (-1.85, 2.85) {$r_2^2\in\mathbb{R}^-$};

\node[font=\tiny] at (-0.1,-0.1) {0};

\draw[->] (-3,0)--(3,0);
\draw[->] (0,-4)--(0,4);

\draw(2,3)--(2,2.5)--(2.5,2.5);
\node[ rotate=0] at (2.3, 2.8) {$l^1$};

\end{tikzpicture}
\caption{Branch points and branch cuts of $\B_{\udon2}$}
\label{fig:Branch points and branch cuts of Bu12}
\end{figure}

\paragraph{The integration}
Since $\eta_1+\eta_2>0$, when $l^1\to \infty$, we have
\begin{equation}
    r_\udon\to -|l^1|, \quad r_2\to|l^1|,
\end{equation}
and
\begin{equation}
    F_\uot\to -i(\eta_1+\eta_2)\frac{l^1}{\vpnorm}, \quad Q_\uot\to -4i(\eta_1+\eta_2){l^1}{\vpnorm}
\end{equation}
Thus,
\begin{equation}\label{eq:betau12 limit}
\begin{aligned}
    \beta_{\udon2}&\to-\frac{2  \pi i|\vp\,|(l^1)^2}{ Q_{\udon2}}\Big(
    \frac{1}{r_\udon q_\udon}- \frac{1}{r_2 q_2}
    \Big)\\&
    \to\frac{2  \pi i|\vp\,|(l^1)^2}{ Q_{\udon2}|l^1|}
    \frac{q_\udon+q_2}{ q_\udon q_2}\\&
    \to-\frac{ \pi }{ 2(\eta_1+\eta_2)}
    \frac{q_\udon+q_2}{ q_\udon q_2}\Sign(l^1).
\end{aligned}
\end{equation}
Then, we have
\begin{equation}
    \int \B_{\udon2}\diff l^l=\lim_{l^1\to \infty}\beta_{\udon2}-\lim_{l^1\to -\infty}\beta_{\udon2}=-\frac{ \pi }{ \eta_1+\eta_2}
    \frac{q_\udon+q_2}{ q_\udon q_2}.
\end{equation}
It is \eqref{eq:IntBu12 definite}.
\subsection{Equation \eqref{eq:I1120 1,3-cut direct integration}}\label{app sec: 2L}

We compute
\begin{equation}
\begin{aligned}
    \I_{1,1,2,0}|_{\Cut{1,3}}=&-\int \frac{\diff^D l_1\diff^D l_2}{(i\pi^{\frac{D}{2}})^2}\frac{\delta(P_1)\delta^\prime(P_3)}{D_2}
\end{aligned}
\end{equation}
Notice that $P_1$ and $D_2$ are free of $l_2$, and $P_3$ is free of $l_1$, the expression is factorized as
\begin{equation}
\begin{aligned}
    \I_{1,1,2,0}|_{\Cut{1,3}}=\Big(2\pi i\int \frac{\diff^D l_1}{i\pi^{\frac{D}{2}}}\frac{\delta(P_1)}{D_2}\Big)\Big(-2\pi i\int \frac{\diff^D l_2}{i\pi^{\frac{D}{2}}}\delta^\prime(P_3)\Big)
\end{aligned}
\end{equation}
For the integration over $l_2$ in the second bracket, it is the same as what we computed in \eqref{eq:1-cut spec calculation} (with $m^2$ replaced by $s$). We can directly write it down as
\begin{equation}
\begin{aligned}
   -2\pi i\int \frac{\diff^D l_2}{i\pi^{\frac{D}{2}}}\delta^\prime(P_3)=-2s^{\frac{D-4}{2}}\Gamma(2-\frac{D}{2})
\end{aligned}
\end{equation}
We now compute the integrate over $l_1$ in the first bracket. It is 
\begin{equation}\label{eq:2L int over l_1^0}
\begin{aligned}
    &2\pi i\int \frac{\diff^{D} l_1}{i\pi^{\frac{D}{2}}}\frac{\delta(P_1)}{D_2}\\
    =&2\pi i\int \frac{\diff^{D-1} \vl_1\diff l_1^0}{i\pi^{\frac{D}{2}}}\frac{\delta((l_1^0)^2-|\vl_1|^2)}{(\sqrt{s}+l_1^0)^2 -|\vl_1|^2-m^2-i \eta_2}\\
    =&2\pi i\int \frac{\diff^{D-1} \vl_1}{i\pi^{\frac{D}{2}}}\frac{1}{2|\vl_1|}\Big(\frac{1}{(\sqrt{s}+|\vl_1|)^2 -|\vl_1|^2-m^2-i \eta_2}+\frac{1}{(\sqrt{s}-|\vl_1|)^2 -|\vl_1|^2-m^2-i \eta_2}\Big)\\
    =&2\pi i \int_0^{+\infty} \frac{\diff|\vl_1||\vl_1|^{D-3}\Omega_{D-1}}{i\pi^{\frac{D}{2}}}\frac{s-m^2-i \eta_2}{(s-m^2-i \eta_2)^2-4 s |\vl_1|^2}\\
    =&\frac{2^{3-D}\pi^{\frac{3}{2}}s^{1-\frac{D}{2}}\csc{(\frac{\pi D}{2})} }{\kappa^3 \Gamma(\frac{D-1}{2})}
\end{aligned}
\end{equation}
where
\begin{equation}
    \kappa:=s-m^2-i\eta_2.
\end{equation}
In the derivations in \eqref{eq:2L int over l_1^0}, we have set
\begin{equation}
    p=(\sqrt{s},\vec{0}^{(D-1)}).
\end{equation}

With the above, we have
\begin{equation}\label{eq:2L direct integration before eta2->0}
    \I_{1,1,2,0}|_{\Cut{1,3}}=\frac{4\pi (i\kappa)^D\cot{(\frac{\pi D}{2})}\Gamma(3-D)}{s\kappa^3}.
\end{equation}
Taking $\eta_2\to\zp$ we get \eqref{eq:I1120 1,3-cut direct integration}.

\section{Checking the equations with {\texttt{Mathematica}}}

The correctness of some equations  (or statements) in this paper are checked using \texttt{Mathematica}. We provide the codes checking the equations in \texttt{Derivations-1.wl} in the GitHub repo.

\begin{center}
    \url{https://github.com/Wu-Zihao/IBP-spc-consistency-problem-studies}
\end{center}

Table \ref{tab:code check} shows the correspondence of the sections in the code and the equations they check.

\begin{table}[hbtp]
    \centering
    \begin{tabular}{|c|c||c|c|}
    \hline
        Code section & Equation(s) it checks&
        Code section & Equation(s) it checks
        
        \\\hline
        
         2.2.1&\eqref{eq:1-cut spec calculation} &
         
         2.2.2&\eqref{eq:B12}
        
         \\\hline

         2.2.3&\eqref{eq:1-cut spec calculation} &
         
         2.2.4&\eqref{eq:B12}
        
         \\\hline

         2.2.5&\eqref{eq:B12 limit} &
         
         2.2.6&\eqref{eq:IntB12 indefinite}, \eqref{eq:rep for beta12}

         \\\hline
         
         2.2.7&\eqref{eq:A5 Q12} &
         
         2.2.8&\eqref{eq:solution Q12=0}

         \\\hline
         
         2.2.9&\eqref{eq:r1 substituted by solution of Q12=0}, \eqref{eq:r2 substituted by solution of Q12=0} &
         
         2.2.10&\eqref{eq:p^2-(r1-r2)^2=0 when Q12=0}

         \\\hline
         
         2.2.11&\eqref{eq:branch cuts r1}, \eqref{eq:branch cuts r2}, \eqref{eq:Ki definition}&
         
         2.2.12&\eqref{eq:beta12 limit}, \eqref{eq:beta12 factor in bracket when eta1=eta2}, \eqref{eq:beta12 limit when eta1=eta2}

         \\\hline
         
         2.2.13&\eqref{eq:Bu12} &
         
         2.2.14&\eqref{eq: Bu12 limit}

         \\\hline
         
         2.2.15&\eqref{eq:IntBu12 indefinite}, \eqref{eq:rep for betau12} &
         
         2.2.16&\eqref{eq:A Qu12}

         \\\hline
         
         2.2.17&\eqref{eq:solution Qu12=0} &
         
         2.2.18&\eqref{eq:ru1 substituted by solution of Qu12=0}, \eqref{eq:r2 substituted by solution of Qu12=0}

         \\\hline

         2.2.19&\eqref{eq:p^2-(r1-r2)^2=0 when Qu12=0} &
         
         2.2.20&\eqref{eq:branch cuts ru1}, \eqref{eq:Ku1 definition}

         \\\hline
         
         2.2.21&\eqref{eq:betau12 limit} &
         
         3.2&Table \ref{tab: massive dbox sectors with hidden relations}

         \\\hline
         
         4.2&\eqref{eq:2L IBP with h.t.} &
         
         4.3&\eqref{eq:2L int over l_1^0}, \eqref{eq:2L direct integration before eta2->0}

         \\\hline
         
         4.4&Remark \ref{remark 1} &
         
         &

         \\\hline
    \end{tabular}
    \caption{The equations checked in the codes}
    \label{tab:code check}
\end{table}

\bibliographystyle{elsarticle-num}
\bibliography{bibtex}

@article{Gluza:2010ws,
    author = "Gluza, Janusz and Kajda, Krzysztof and Kosower, David A.",
    title = "{Towards a Basis for Planar Two-Loop Integrals}",
    eprint = "1009.0472",
    archivePrefix = "arXiv",
    primaryClass = "hep-th",
    reportNumber = "SACLAY-IPHT-T10-089, WIS-09-10-JULY-DPPA",
    doi = "10.1103/PhysRevD.83.045012",
    journal = "Phys. Rev. D",
    volume = "83",
    pages = "045012",
    year = "2011"
}

@article{Larsen:2015ped,
    author = "Larsen, Kasper J. and Zhang, Yang",
    title = "{Integration-by-parts reductions from unitarity cuts and algebraic geometry}",
    eprint = "1511.01071",
    archivePrefix = "arXiv",
    primaryClass = "hep-th",
    doi = "10.1103/PhysRevD.93.041701",
    journal = "Phys. Rev. D",
    volume = "93",
    number = "4",
    pages = "041701",
    year = "2016"
}

@inproceedings{Zhang:2016kfo,
    author = "Zhang, Yang",
    title = "{Lecture Notes on Multi-loop Integral Reduction and Applied Algebraic Geometry}",
    eprint = "1612.02249",
    archivePrefix = "arXiv",
    primaryClass = "hep-th",
    month = "12",
    year = "2016"
}

@article{Schabinger:2011dz,
    author = "Schabinger, Robert M.",
    title = "{A New Algorithm For The Generation Of Unitarity-Compatible Integration By Parts Relations}",
    eprint = "1111.4220",
    archivePrefix = "arXiv",
    primaryClass = "hep-ph",
    reportNumber = "IFT-UAM-CSIC-11-89",
    doi = "10.1007/JHEP01(2012)077",
    journal = "JHEP",
    volume = "01",
    pages = "077",
    year = "2012"
}

@article{Bohm:2017qme,
    author = {B\"ohm, Janko and Georgoudis, Alessandro and Larsen, Kasper J. and Schulze, Mathias and Zhang, Yang},
    title = "{Complete sets of logarithmic vector fields for integration-by-parts identities of Feynman integrals}",
    eprint = "1712.09737",
    archivePrefix = "arXiv",
    primaryClass = "hep-th",
    reportNumber = "MITP-17-104, UUITP-44-17, MITP/17-104, UUITP-44/17",
    doi = "10.1103/PhysRevD.98.025023",
    journal = "Phys. Rev. D",
    volume = "98",
    number = "2",
    pages = "025023",
    year = "2018"
}

@article{Boehm:2020zig,
    author = "Boehm, Janko and Bendle, Dominik and Decker, Wolfram and Georgoudis, Alessandro and Pfreundt, Franz-Josef and Rahn, Mirko and Zhang, Yang",
    title = "{Module Intersection for the Integration-by-Parts Reduction of Multi-Loop Feynman Integrals}",
    eprint = "2010.06895",
    archivePrefix = "arXiv",
    primaryClass = "hep-th",
    doi = "10.22323/1.383.0004",
    journal = "PoS",
    volume = "MA2019",
    pages = "004",
    year = "2022"
}

@misc{Mathematica,
  author = {Wolfram Research{,} Inc.},
  title = {Mathematica, {V}ersion 13.2},
  url = {https://www.wolfram.com/mathematica},
  note = {Champaign, IL, 2022}
}

@article{Peraro:2019svx,
    author = "Peraro, Tiziano",
    title = "{FiniteFlow: multivariate functional reconstruction using finite fields and dataflow graphs}",
    eprint = "1905.08019",
    archivePrefix = "arXiv",
    primaryClass = "hep-ph",
    reportNumber = "ZU-TH 24/19",
    doi = "10.1007/JHEP07(2019)031",
    journal = "JHEP",
    volume = "07",
    pages = "031",
    year = "2019"
}

@article{Lee:2013mka,
    author = "Lee, Roman N.",
    editor = "Wang, Jianxiong",
    title = "{LiteRed 1.4: a powerful tool for reduction of multiloop integrals}",
    eprint = "1310.1145",
    archivePrefix = "arXiv",
    primaryClass = "hep-ph",
    doi = "10.1088/1742-6596/523/1/012059",
    journal = "J. Phys. Conf. Ser.",
    volume = "523",
    pages = "012059",
    year = "2014"
}

@article{Lee:2013hzt,
    author = "Lee, Roman N. and Pomeransky, Andrei A.",
    title = "{Critical points and number of master integrals}",
    eprint = "1308.6676",
    archivePrefix = "arXiv",
    primaryClass = "hep-ph",
    doi = "10.1007/JHEP11(2013)165",
    journal = "JHEP",
    volume = "11",
    pages = "165",
    year = "2013"
}

@article{Lee:2008tj,
    author = "Lee, R. N.",
    title = "{Group structure of the integration-by-part identities and its application to the reduction of multiloop integrals}",
    eprint = "0804.3008",
    archivePrefix = "arXiv",
    primaryClass = "hep-ph",
    doi = "10.1088/1126-6708/2008/07/031",
    journal = "JHEP",
    volume = "07",
    pages = "031",
    year = "2008"
}

@article{Boehm:2020ijp,
    author = "Boehm, Janko and Wittmann, Marcel and Wu, Zihao and Xu, Yingxuan and Zhang, Yang",
    title = "{IBP reduction coefficients made simple}",
    eprint = "2008.13194",
    archivePrefix = "arXiv",
    primaryClass = "hep-ph",
    reportNumber = "MPP-2020-139, PCFT-20-26, USTC-ICTS-20-26",
    doi = "10.1007/JHEP12(2020)054",
    journal = "JHEP",
    volume = "12",
    pages = "054",
    year = "2020"
}

@article{Frellesvig:2020qot,
    author = "Frellesvig, Hjalte and Gasparotto, Federico and Laporta, Stefano and Mandal, Manoj K. and Mastrolia, Pierpaolo and Mattiazzi, Luca and Mizera, Sebastian",
    title = "{Decomposition of Feynman Integrals by Multivariate Intersection Numbers}",
    eprint = "2008.04823",
    archivePrefix = "arXiv",
    primaryClass = "hep-th",
    doi = "10.1007/JHEP03(2021)027",
    journal = "JHEP",
    volume = "03",
    pages = "027",
    year = "2021"
}

@article{Smirnov:2019qkx,
    author = "Smirnov, A. V. and Chuharev, F. S.",
    title = "{FIRE6: Feynman Integral REduction with Modular Arithmetic}",
    eprint = "1901.07808",
    archivePrefix = "arXiv",
    primaryClass = "hep-ph",
    reportNumber = "TTP19-006",
    doi = "10.1016/j.cpc.2019.106877",
    journal = "Comput. Phys. Commun.",
    volume = "247 ",
    pages = "106877",
    year = "2020"
}

@article{Liu:2021wks,
    author = "Liu, Xiao and Ma, Yan-Qing",
    title = "{Multiloop corrections for collider processes using auxiliary mass flow}",
    eprint = "2107.01864",
    archivePrefix = "arXiv",
    primaryClass = "hep-ph",
    doi = "10.1103/PhysRevD.105.L051503",
    journal = "Phys. Rev. D",
    volume = "105",
    number = "5",
    pages = "L051503",
    year = "2022"
}

@article{Liu:2018dmc,
    author = "Liu, Xiao and Ma, Yan-Qing",
    title = "{Determining arbitrary Feynman integrals by vacuum integrals}",
    eprint = "1801.10523",
    archivePrefix = "arXiv",
    primaryClass = "hep-ph",
    doi = "10.1103/PhysRevD.99.071501",
    journal = "Phys. Rev. D",
    volume = "99",
    number = "7",
    pages = "071501",
    year = "2019"
}

@article{Bendle:2019csk,
    author = {Bendle, Dominik and B\"ohm, Janko and Decker, Wolfram and Georgoudis, Alessandro and Pfreundt, Franz-Josef and Rahn, Mirko and Wasser, Pascal and Zhang, Yang},
    title = "{Integration-by-parts reductions of Feynman integrals using Singular and GPI-Space}",
    eprint = "1908.04301",
    archivePrefix = "arXiv",
    primaryClass = "hep-th",
    reportNumber = "MITP/19-055,MPP-2019-164,UUITP-30/19,USTC-ICTS-19-20, MITP/19-055,UUITP-30/19,USTC-ICTS-19-20",
    doi = "10.1007/JHEP02(2020)079",
    journal = "JHEP",
    volume = "02",
    pages = "079",
    year = "2020"
}

@article{Tkachov:1981wb,
    author = "Tkachov, F. V.",
    title = "{A Theorem on Analytical Calculability of Four Loop Renormalization Group Functions}",
    doi = "10.1016/0370-2693(81)90288-4",
    journal = "Phys. Lett. B",
    volume = "100",
    pages = "65--68",
    year = "1981"
}

@article{Chetyrkin:1981qh,
    author = "Chetyrkin, K. G. and Tkachov, F. V.",
    title = "{Integration by Parts: The Algorithm to Calculate beta Functions in 4 Loops}",
    doi = "10.1016/0550-3213(81)90199-1",
    journal = "Nucl. Phys. B",
    volume = "192",
    pages = "159--204",
    year = "1981"
}

@article{Smirnov:2006wh,
    author = "Smirnov, A. V. and Smirnov, V. A.",
    editor = "Blumlein, J. and Moch, S. and Riemann, T.",
    title = "{S-bases as a tool to solve reduction problems for Feynman integrals}",
    eprint = "hep-ph/0606247",
    archivePrefix = "arXiv",
    doi = "10.1016/j.nuclphysbps.2006.09.032",
    journal = "Nucl. Phys. B Proc. Suppl.",
    volume = "160",
    pages = "80--84",
    year = "2006"
}

@article{Smirnov:2006tz,
    author = "Smirnov, A. V.",
    title = "{An Algorithm to construct Grobner bases for solving integration by parts relations}",
    eprint = "hep-ph/0602078",
    archivePrefix = "arXiv",
    doi = "10.1088/1126-6708/2006/04/026",
    journal = "JHEP",
    volume = "04",
    pages = "026",
    year = "2006"
}

@article{Ita:2015tya,
    author = "Ita, Harald",
    title = "{Two-loop Integrand Decomposition into Master Integrals and Surface Terms}",
    eprint = "1510.05626",
    archivePrefix = "arXiv",
    primaryClass = "hep-th",
    reportNumber = "FR-PHENO-2015-011",
    doi = "10.1103/PhysRevD.94.116015",
    journal = "Phys. Rev. D",
    volume = "94",
    number = "11",
    pages = "116015",
    year = "2016"
}

@article{vonManteuffel:2020vjv,
    author = "von Manteuffel, Andreas and Panzer, Erik and Schabinger, Robert M.",
    title = "{Cusp and collinear anomalous dimensions in four-loop QCD from form factors}",
    eprint = "2002.04617",
    archivePrefix = "arXiv",
    primaryClass = "hep-ph",
    reportNumber = "MSUHEP-20-002",
    doi = "10.1103/PhysRevLett.124.162001",
    journal = "Phys. Rev. Lett.",
    volume = "124",
    number = "16",
    pages = "162001",
    year = "2020"
}

@article{Mastrolia:2018uzb,
    author = "Mastrolia, Pierpaolo and Mizera, Sebastian",
    title = "{Feynman Integrals and Intersection Theory}",
    eprint = "1810.03818",
    archivePrefix = "arXiv",
    primaryClass = "hep-th",
    doi = "10.1007/JHEP02(2019)139",
    journal = "JHEP",
    volume = "02",
    pages = "139",
    year = "2019"
}

@article{Frellesvig:2019uqt,
    author = "Frellesvig, Hjalte and Gasparotto, Federico and Mandal, Manoj K. and Mastrolia, Pierpaolo and Mattiazzi, Luca and Mizera, Sebastian",
    title = "{Vector Space of Feynman Integrals and Multivariate Intersection Numbers}",
    eprint = "1907.02000",
    archivePrefix = "arXiv",
    primaryClass = "hep-th",
    doi = "10.1103/PhysRevLett.123.201602",
    journal = "Phys. Rev. Lett.",
    volume = "123",
    number = "20",
    pages = "201602",
    year = "2019"
}

@article{Frellesvig:2019kgj,
    author = "Frellesvig, Hjalte and Gasparotto, Federico and Laporta, Stefano and Mandal, Manoj K. and Mastrolia, Pierpaolo and Mattiazzi, Luca and Mizera, Sebastian",
    title = "{Decomposition of Feynman Integrals on the Maximal Cut by Intersection Numbers}",
    eprint = "1901.11510",
    archivePrefix = "arXiv",
    primaryClass = "hep-ph",
    doi = "10.1007/JHEP05(2019)153",
    journal = "JHEP",
    volume = "05",
    pages = "153",
    year = "2019"
}

@article{Guan:2019bcx,
    author = "Guan, Xin and Liu, Xiao and Ma, Yan-Qing",
    title = "{Complete reduction of integrals in two-loop five-light-parton scattering amplitudes}",
    eprint = "1912.09294",
    archivePrefix = "arXiv",
    primaryClass = "hep-ph",
    doi = "10.1088/1674-1137/44/9/093106",
    journal = "Chin. Phys. C",
    volume = "44",
    number = "9",
    pages = "093106",
    year = "2020"
}

@article{vonManteuffel:2014ixa,
    author = "von Manteuffel, Andreas and Schabinger, Robert M.",
    title = "{A novel approach to integration by parts reduction}",
    eprint = "1406.4513",
    archivePrefix = "arXiv",
    primaryClass = "hep-ph",
    reportNumber = "MITP-14-009",
    doi = "10.1016/j.physletb.2015.03.029",
    journal = "Phys. Lett. B",
    volume = "744",
    pages = "101--104",
    year = "2015"
}

@article{Peraro:2016wsq,
    author = "Peraro, Tiziano",
    title = "{Scattering amplitudes over finite fields and multivariate functional reconstruction}",
    eprint = "1608.01902",
    archivePrefix = "arXiv",
    primaryClass = "hep-ph",
    reportNumber = "EDINBURGH-2016-14",
    doi = "10.1007/JHEP12(2016)030",
    journal = "JHEP",
    volume = "12",
    pages = "030",
    year = "2016"
}

@article{Klappert:2019emp,
    author = "Klappert, Jonas and Lange, Fabian",
    title = "{Reconstructing rational functions with FireFly}",
    eprint = "1904.00009",
    archivePrefix = "arXiv",
    primaryClass = "cs.SC",
    reportNumber = "TTK-19-11, P3H-19-007",
    doi = "10.1016/j.cpc.2019.106951",
    journal = "Comput. Phys. Commun.",
    volume = "247",
    pages = "106951",
    year = "2020"
}

@article{Kosower:2018obg,
    author = "Kosower, David A.",
    title = "{Direct Solution of Integration-by-Parts Systems}",
    eprint = "1804.00131",
    archivePrefix = "arXiv",
    primaryClass = "hep-ph",
    doi = "10.1103/PhysRevD.98.025008",
    journal = "Phys. Rev. D",
    volume = "98",
    number = "2",
    pages = "025008",
    year = "2018"
}

@article{Anastasiou:2004vj,
    author = "Anastasiou, Charalampos and Lazopoulos, Achilleas",
    title = "{Automatic integral reduction for higher order perturbative calculations}",
    eprint = "hep-ph/0404258",
    archivePrefix = "arXiv",
    reportNumber = "SLAC-PUB-10428",
    doi = "10.1088/1126-6708/2004/07/046",
    journal = "JHEP",
    volume = "07",
    pages = "046",
    year = "2004"
}

@article{Smirnov:2008iw,
    author = "Smirnov, A. V.",
    title = "{Algorithm FIRE -- Feynman Integral REduction}",
    eprint = "0807.3243",
    archivePrefix = "arXiv",
    primaryClass = "hep-ph",
    reportNumber = "TTP08-30, SFB-CPP-08-51",
    doi = "10.1088/1126-6708/2008/10/107",
    journal = "JHEP",
    volume = "10",
    pages = "107",
    year = "2008"
}

@article{Smirnov:2013dia,
    author = "Smirnov, A. V. and Smirnov, V. A.",
    title = "{FIRE4, LiteRed and accompanying tools to solve integration by parts relations}",
    eprint = "1302.5885",
    archivePrefix = "arXiv",
    primaryClass = "hep-ph",
    reportNumber = "HU-EP-13-04, HU-MATHEMATIK:05-2013",
    doi = "10.1016/j.cpc.2013.06.016",
    journal = "Comput. Phys. Commun.",
    volume = "184",
    pages = "2820--2827",
    year = "2013"
}

@article{Smirnov:2014hma,
    author = "Smirnov, Alexander V.",
    title = "{FIRE5: a C++ implementation of Feynman Integral REduction}",
    eprint = "1408.2372",
    archivePrefix = "arXiv",
    primaryClass = "hep-ph",
    reportNumber = "SFB-CPP-14-60",
    doi = "10.1016/j.cpc.2014.11.024",
    journal = "Comput. Phys. Commun.",
    volume = "189",
    pages = "182--191",
    year = "2015"
}

@article{Maierhofer:2017gsa,
    author = {Maierh\"ofer, Philipp and Usovitsch, Johann and Uwer, Peter},
    title = "{Kira\textemdash{}A Feynman integral reduction program}",
    eprint = "1705.05610",
    archivePrefix = "arXiv",
    primaryClass = "hep-ph",
    doi = "10.1016/j.cpc.2018.04.012",
    journal = "Comput. Phys. Commun.",
    volume = "230",
    pages = "99--112",
    year = "2018"
}

@article{Maierhofer:2018gpa,
    author = {Maierh\"ofer, Philipp and Usovitsch, Johann},
    title = "{Kira 1.2 Release Notes}",
    eprint = "1812.01491",
    archivePrefix = "arXiv",
    primaryClass = "hep-ph",
    reportNumber = "FR-PHENO-2018-15, TCDMATH 18-18",
    month = "12",
    year = "2018"
}

@article{Maierhofer:2019goc,
    author = {Maierh\"ofer, P. and Usovitsch, J.},
    editor = "Blondel, A. and Gluza, J. and Jadach, S. and Janot, P. and Riemann, T.",
    title = "{Recent developments in Kira}",
    doi = "10.23731/CYRM-2020-003.201",
    journal = "CERN Yellow Reports: Monographs",
    volume = "3",
    pages = "201--204",
    year = "2020"
}

@article{Studerus:2009ye,
    author = "Studerus, C.",
    title = "{Reduze-Feynman Integral Reduction in C++}",
    eprint = "0912.2546",
    archivePrefix = "arXiv",
    primaryClass = "physics.comp-ph",
    reportNumber = "ZU-TH-18-09",
    doi = "10.1016/j.cpc.2010.03.012",
    journal = "Comput. Phys. Commun.",
    volume = "181",
    pages = "1293--1300",
    year = "2010"
}

@article{vonManteuffel:2012np,
    author = "von Manteuffel, A. and Studerus, C.",
    title = "{Reduze 2 - Distributed Feynman Integral Reduction}",
    eprint = "1201.4330",
    archivePrefix = "arXiv",
    primaryClass = "hep-ph",
    reportNumber = "ZU-TH-01-12, MZ-TH-12-03, BI-TP-2012-02",
    month = "1",
    year = "2012"
}

@article{Baikov:1996cd,
    author = "Baikov, P. A.",
    title = "{Explicit solutions of n loop vacuum integral recurrence relations}",
    eprint = "hep-ph/9604254",
    archivePrefix = "arXiv",
    reportNumber = "INP-96-10-418",
    month = "4",
    year = "1996"
}

@article{Baikov:1996rk,
    author = "Baikov, P. A.",
    title = "{Explicit solutions of the three loop vacuum integral recurrence relations}",
    eprint = "hep-ph/9603267",
    archivePrefix = "arXiv",
    reportNumber = "INP-96-10-417",
    doi = "10.1016/0370-2693(96)00835-0",
    journal = "Phys. Lett. B",
    volume = "385",
    pages = "404--410",
    year = "1996"
}

@article{Baikov:2005nv,
    author = "Baikov, P. A.",
    title = "{A Practical criterion of irreducibility of multi-loop Feynman integrals}",
    eprint = "hep-ph/0507053",
    archivePrefix = "arXiv",
    doi = "10.1016/j.physletb.2006.01.052",
    journal = "Phys. Lett. B",
    volume = "634",
    pages = "325--329",
    year = "2006"
}

@article{Klappert:2020aqs,
    author = "Klappert, Jonas and Klein, Sven Yannick and Lange, Fabian",
    title = "{Interpolation of dense and sparse rational functions and other improvements in FireFly}",
    eprint = "2004.01463",
    archivePrefix = "arXiv",
    primaryClass = "cs.MS",
    reportNumber = "TTK-20-07, P3H-20-010",
    doi = "10.1016/j.cpc.2021.107968",
    journal = "Comput. Phys. Commun.",
    volume = "264",
    pages = "107968",
    year = "2021"
}

@article{Georgoudis:2016wff,
    author = "Georgoudis, Alessandro and Larsen, Kasper J. and Zhang, Yang",
    title = "{Azurite: An algebraic geometry based package for finding bases of loop integrals}",
    eprint = "1612.04252",
    archivePrefix = "arXiv",
    primaryClass = "hep-th",
    doi = "10.1016/j.cpc.2017.08.013",
    journal = "Comput. Phys. Commun.",
    volume = "221",
    pages = "203--215",
    year = "2017"
}

@inproceedings{Lee:2014tja,
    author = "Lee, Roman N.",
    title = "{Modern techniques of multiloop calculations}",
    booktitle = "{49th Rencontres de Moriond on QCD and High Energy Interactions}",
    eprint = "1405.5616",
    archivePrefix = "arXiv",
    primaryClass = "hep-ph",
    pages = "297--300",
    year = "2014"
}

@article{Smirnov:2006vt,
    author = "Smirnov, Vladimir A.",
    editor = "Fujimoto, J. and Kodaira, J. and Uematsu, T.",
    title = "{Some recent results on evaluating Feynman integrals}",
    eprint = "hep-ph/0601268",
    archivePrefix = "arXiv",
    doi = "10.1016/j.nuclphysbps.2006.03.017",
    journal = "Nucl. Phys. B Proc. Suppl.",
    volume = "157",
    pages = "131--135",
    year = "2006"
}

@article{Smirnov:2005ky,
    author = "Smirnov, A. V. and Smirnov, Vladimir A.",
    title = "{Applying Grobner bases to solve reduction problems for Feynman integrals}",
    eprint = "hep-lat/0509187",
    archivePrefix = "arXiv",
    doi = "10.1088/1126-6708/2006/01/001",
    journal = "JHEP",
    volume = "01",
    pages = "001",
    year = "2006"
}

@article{Bitoun:2017nre,
    author = "Bitoun, Thomas and Bogner, Christian and Klausen, Rene Pascal and Panzer, Erik",
    title = "{Feynman integral relations from parametric annihilators}",
    eprint = "1712.09215",
    archivePrefix = "arXiv",
    primaryClass = "hep-th",
    doi = "10.1007/s11005-018-1114-8",
    journal = "Lett. Math. Phys.",
    volume = "109",
    number = "3",
    pages = "497--564",
    year = "2019"
}

@article{Baikov:1996iu,
    author = "Baikov, P. A.",
    editor = "Werlen, M. and Perret-Gallix, D.",
    title = "{Explicit solutions of the multiloop integral recurrence relations and its application}",
    eprint = "hep-ph/9611449",
    archivePrefix = "arXiv",
    reportNumber = "INP-96-42-449",
    doi = "10.1016/S0168-9002(97)00126-5",
    journal = "Nucl. Instrum. Meth. A",
    volume = "389",
    pages = "347--349",
    year = "1997"
}

@article{Grozin:2011mt,
    author = "Grozin, A. G.",
    title = "{Integration by parts: An Introduction}",
    eprint = "1104.3993",
    archivePrefix = "arXiv",
    primaryClass = "hep-ph",
    reportNumber = "TTP11-09",
    doi = "10.1142/S0217751X11053687",
    journal = "Int. J. Mod. Phys. A",
    volume = "26",
    pages = "2807--2854",
    year = "2011"
}

@article{Smirnov_2006,
doi = {10.1088/1126-6708/2006/01/001},
url = {https://doi.org/10.1088\%2F1126-6708\%2F2006\%2F01\%2F001},
year = 2006,
month = {jan},
publisher = {Springer Science and Business Media {LLC}
},
volume = {2006},
number = {01},
pages = {001--001},
author = {Alexander V Smirnov and Vladimir A Smirnov},
title = {Applying Gröbner bases to solve reduction problems for Feynman integrals},
journal = {Journal of High Energy Physics}
}

@book{Smirnov:2012gma,
    author = "Smirnov, Vladimir A.",
    title = "{Analytic tools for Feynman integrals}",
    doi = "10.1007/978-3-642-34886-0",
    volume = "250",
    year = "2012"
}

@article{Wu:2023upw,
    author = "Wu, Zihao and Boehm, Janko and Ma, Rourou and Xu, Hefeng and Zhang, Yang",
    title = "{NeatIBP 1.0, A package generating small-size integration-by-parts relations for Feynman integrals}",
    eprint = "2305.08783",
    archivePrefix = "arXiv",
    primaryClass = "hep-ph",
    reportNumber = "USTC-ICTS/PCFT-23-15",
    month = "5",
    year = "2023"
}

@article{Guan:2024byi,
    author = "Guan, Xin and Liu, Xiao and Ma, Yan-Qing and Wu, Wen-Hao",
    title = "{Blade: A package for block-triangular form improved Feynman integrals decomposition}",
    eprint = "2405.14621",
    archivePrefix = "arXiv",
    primaryClass = "hep-ph",
    doi = "10.1016/j.cpc.2025.109538",
    journal = "Comput. Phys. Commun.",
    volume = "310",
    pages = "109538",
    year = "2025"
}

@article{Klappert:2020nbg,
    author = {Klappert, Jonas and Lange, Fabian and Maierh\"ofer, Philipp and Usovitsch, Johann},
    title = "{Integral reduction with Kira 2.0 and finite field methods}",
    eprint = "2008.06494",
    archivePrefix = "arXiv",
    primaryClass = "hep-ph",
    reportNumber = "TTK-20-24, P3H-20-041, FR-PHENO-2020-11, MITP/20-044",
    doi = "10.1016/j.cpc.2021.108024",
    journal = "Comput. Phys. Commun.",
    volume = "266",
    pages = "108024",
    year = "2021"
}

@article{Lange:2024mmz,
    author = "Lange, Fabian and Usovitsch, Johann and Wu, Zihao",
    title = "{Towards the next Kira release}",
    eprint = "2407.01395",
    archivePrefix = "arXiv",
    primaryClass = "hep-ph",
    reportNumber = "ZU-TH 32/24, PSI-PR-24-14, CERN-TH-2024-096",
    doi = "10.22323/1.467.0070",
    journal = "PoS",
    volume = "LL2024",
    pages = "070",
    year = "2024"
}

@article{Usovitsch:2022wvr,
    author = "Usovitsch, Johann",
    title = "{Kira and the block-triangular form}",
    doi = "10.22323/1.416.0071",
    journal = "PoS",
    volume = "LL2022",
    pages = "071",
    year = "2022"
}

@article{Wu:2025aeg,
    author = {Wu, Zihao and B\"ohm, Janko and Ma, Rourou and Usovitsch, Johann and Xu, Yingxuan and Zhang, Yang},
    title = "{Performing integration-by-parts reductions using NeatIBP 1.1 + Kira}",
    eprint = "2502.20778",
    archivePrefix = "arXiv",
    primaryClass = "hep-ph",
    reportNumber = "USTC-ICTS/PCFT-25-10, MPP-2025-29, HU-EP-25/12-RTG",
    month = "2",
    year = "2025"
}

@article{Lange:2025fba,
    author = "Lange, Fabian and Usovitsch, Johann and Wu, Zihao",
    title = "{Kira 3: integral reduction with efficient seeding and optimized equation selection}",
    eprint = "2505.20197",
    archivePrefix = "arXiv",
    primaryClass = "hep-ph",
    reportNumber = "ZU-TH 39/25, HU-EP-25/17-RTG",
    month = "5",
    year = "2025"
}

@article{Smirnov:2023yhb,
    author = "Smirnov, Alexander V. and Zeng, Mao",
    title = "{FIRE 6.5: Feynman integral reduction with new simplification library}",
    eprint = "2311.02370",
    archivePrefix = "arXiv",
    primaryClass = "hep-ph",
    doi = "10.1016/j.cpc.2024.109261",
    journal = "Comput. Phys. Commun.",
    volume = "302",
    pages = "109261",
    year = "2024"
}

@article{Magerya:2022hvj,
    author = "Magerya, Vitaly",
    title = "{Rational Tracer: a Tool for Faster Rational Function Reconstruction}",
    eprint = "2211.03572",
    archivePrefix = "arXiv",
    primaryClass = "physics.data-an",
    reportNumber = "KA-TP-26-2022, P3H-22-109",
    month = "11",
    year = "2022"
}

@article{Zeng:2025xbh,
    author = "Zeng, Mao",
    title = "{Reinforcement Learning and Metaheuristics for Feynman Integral Reduction}",
    eprint = "2504.16045",
    archivePrefix = "arXiv",
    primaryClass = "hep-ph",
    month = "4",
    year = "2025"
}

@article{Smirnov:2024onl,
    author = "Smirnov, Alexander and Zeng, Mao",
    title = "{Feynman integral reduction: balanced reconstruction of sparse rational functions and implementation on supercomputers in a co-design approach}",
    eprint = "2409.19099",
    archivePrefix = "arXiv",
    primaryClass = "hep-ph",
    doi = "10.26089/NumMet.2024s03",
    month = "9",
    year = "2024"
}

@article{vonHippel:2025okr,
    author = "von Hippel, Matt and Wilhelm, Matthias",
    title = "{Refining Integration-by-Parts Reduction of Feynman Integrals with Machine Learning}",
    eprint = "2502.05121",
    archivePrefix = "arXiv",
    primaryClass = "hep-th",
    doi = "10.1007/JHEP05(2025)185",
    journal = "JHEP",
    volume = "05",
    pages = "185",
    year = "2025"
}

@article{Song:2025pwy,
    author = "Song, Zhuo-Yang and Yang, Tong-Zhi and Cao, Qing-Hong and Luo, Ming-xing and Zhu, Hua Xing",
    title = "{Explainable AI-assisted Optimization for Feynman Integral Reduction}",
    eprint = "2502.09544",
    archivePrefix = "arXiv",
    primaryClass = "hep-ph",
    reportNumber = "ZU-TH 07/25",
    month = "2",
    year = "2025"
}

@article{Britto:2004nc,
    author = "Britto, Ruth and Cachazo, Freddy and Feng, Bo",
    title = "{Generalized unitarity and one-loop amplitudes in N=4 super-Yang-Mills}",
    eprint = "hep-th/0412103",
    archivePrefix = "arXiv",
    doi = "10.1016/j.nuclphysb.2005.07.014",
    journal = "Nucl. Phys. B",
    volume = "725",
    pages = "275--305",
    year = "2005"
}

@article{Bohm:2018bdy,
    author = {B{\"o}hm, Janko and Georgoudis, Alessandro and Larsen, Kasper J. and Sch{\"o}nemann, Hans and Zhang, Yang},
    title = "{Complete integration-by-parts reductions of the non-planar hexagon-box via module intersections}",
    eprint = "1805.01873",
    archivePrefix = "arXiv",
    primaryClass = "hep-th",
    reportNumber = "MITP/18-034, UUITP-16/18, MITP-18-034, UUITP-16-18",
    doi = "10.1007/JHEP09(2018)024",
    journal = "JHEP",
    volume = "09",
    pages = "024",
    year = "2018"
}

@article{Bern:1994zx,
    author = "Bern, Zvi and Dixon, Lance J. and Dunbar, David C. and Kosower, David A.",
    title = "{One loop n point gauge theory amplitudes, unitarity and collinear limits}",
    eprint = "hep-ph/9403226",
    archivePrefix = "arXiv",
    reportNumber = "SLAC-PUB-6415, SACLAY-SPH-T-94-20, UCLA-TEP-94-4, SWAT-94-17",
    doi = "10.1016/0550-3213(94)90179-1",
    journal = "Nucl. Phys. B",
    volume = "425",
    pages = "217--260",
    year = "1994"
}

@article{Bern:1994cg,
    author = "Bern, Zvi and Dixon, Lance J. and Dunbar, David C. and Kosower, David A.",
    title = "{Fusing gauge theory tree amplitudes into loop amplitudes}",
    eprint = "hep-ph/9409265",
    archivePrefix = "arXiv",
    reportNumber = "SLAC-PUB-6563, SACLAY-SPH-T-94-95, UCLA-TEP-94-29, SWAT-94-36",
    doi = "10.1016/0550-3213(94)00488-Z",
    journal = "Nucl. Phys. B",
    volume = "435",
    pages = "59--101",
    year = "1995"
}

@article{Berger:2008sj,
    author = "Berger, C. F. and Bern, Z. and Dixon, L. J. and Febres Cordero, F. and Forde, D. and Ita, H. and Kosower, D. A. and Maitre, D.",
    title = "{An Automated Implementation of On-Shell Methods for One-Loop Amplitudes}",
    eprint = "0803.4180",
    archivePrefix = "arXiv",
    primaryClass = "hep-ph",
    reportNumber = "SLAC-PUB-13161, UCLA-08-TEP-10, MIT-CTP-3937, NSF-KITP-08-48, SACLAY-IPHT-T08-054",
    doi = "10.1103/PhysRevD.78.036003",
    journal = "Phys. Rev. D",
    volume = "78",
    pages = "036003",
    year = "2008"
}

@article{Bern:2004cz,
    author = "Bern, Zvi and Dixon, Lance J. and Kosower, David A.",
    title = "{Two-loop g ---{\ensuremath{>}} gg splitting amplitudes in QCD}",
    eprint = "hep-ph/0404293",
    archivePrefix = "arXiv",
    reportNumber = "SLAC-PUB-10414, UCLA-04-TEP-12, SACLAY-SPHT-T04-051, NSF-KITP-04-43",
    doi = "10.1088/1126-6708/2004/08/012",
    journal = "JHEP",
    volume = "08",
    pages = "012",
    year = "2004"
}

@article{Bern:2000dn,
    author = "Bern, Z. and Dixon, Lance J. and Kosower, D. A.",
    title = "{A Two loop four gluon helicity amplitude in QCD}",
    eprint = "hep-ph/0001001",
    archivePrefix = "arXiv",
    reportNumber = "SLAC-PUB-8324, UCLA-99-TEP-48, SACLAY-SPH-T-99-147",
    doi = "10.1088/1126-6708/2000/01/027",
    journal = "JHEP",
    volume = "01",
    pages = "027",
    year = "2000"
}

@article{Bern:1997sc,
    author = "Bern, Zvi and Dixon, Lance J. and Kosower, David A.",
    title = "{One loop amplitudes for e+ e- to four partons}",
    eprint = "hep-ph/9708239",
    archivePrefix = "arXiv",
    reportNumber = "SLAC-PUB-7529, SACLAY-SPH-T-97-090, UCLA-97-TEP-10",
    doi = "10.1016/S0550-3213(97)00703-7",
    journal = "Nucl. Phys. B",
    volume = "513",
    pages = "3--86",
    year = "1998"
}

@article{Shih:2026jfe,
    author = "Shih, David",
    title = "{Learning to Unscramble Feynman Loop Integrals with SAILIR}",
    eprint = "2604.05034",
    archivePrefix = "arXiv",
    primaryClass = "hep-ph",
    month = "4",
    year = "2026"
}

@article{Huang:2024nij,
    author = "Huang, Li-Hong and Huang, Rui-Jun and Ma, Yan-Qing",
    title = "{Tame multi-leg Feynman integrals beyond one loop}",
    eprint = "2412.21053",
    archivePrefix = "arXiv",
    primaryClass = "hep-ph",
    month = "12",
    year = "2024"
}

@article{Huang:2026xnq,
    author = "Huang, Li-Hong and Ma, Yan-Qing and Wang, Ziwen and Yang, Li Lin",
    title = "{Feynman integral reduction with intersection theory made simple}",
    eprint = "2604.05025",
    archivePrefix = "arXiv",
    primaryClass = "hep-th",
    month = "4",
    year = "2026"
}

@article{Chicherin:2025mvc,
    author = "Chicherin, Dmitry and Wu, Yu and Wu, Zihao and Xu, Yongqun and Zhang, Shun-Qing and Zhang, Yang",
    title = "{Complete computation of all three-loop five-point massless planar integrals}",
    eprint = "2512.17330",
    archivePrefix = "arXiv",
    primaryClass = "hep-ph",
    reportNumber = "MPP-2025-232, USTC-ICTS/PCFT-25-61, LAPTH-056/25",
    month = "12",
    year = "2025"
}

@article{Bern:2024adl,
    author = "Bern, Zvi and Herrmann, Enrico and Roiban, Radu and Ruf, Michael S. and Smirnov, Alexander V. and Smirnov, Vladimir A. and Zeng, Mao",
    title = "{Amplitudes, supersymmetric black hole scattering at $ \mathcal{O}\left({G}^5\right) $, and loop integration}",
    eprint = "2406.01554",
    archivePrefix = "arXiv",
    primaryClass = "hep-th",
    doi = "10.1007/JHEP10(2024)023",
    journal = "JHEP",
    volume = "10",
    pages = "023",
    year = "2024"
}

@article{Larsen_2016,
   title={Integration-by-parts reductions from unitarity cuts and algebraic geometry},
   volume={93},
   ISSN={2470-0029},
   url={http://dx.doi.org/10.1103/PhysRevD.93.041701},
   DOI={10.1103/physrevd.93.041701},
   number={4},
   journal={Physical Review D},
   publisher={American Physical Society (APS)},
   author={Larsen, Kasper J. and Zhang, Yang},
   year={2016},
   month=Feb }

@article{Smirnov:2025prc,
    author = "Smirnov, Alexander V. and Zeng, Mao",
    title = "{FIRE 7: Automatic Reduction with Modular Approach}",
    eprint = "2510.07150",
    archivePrefix = "arXiv",
    primaryClass = "hep-ph",
    month = "10",
    year = "2025"
}

@article{Crisanti:2026rbc,
    author = "Crisanti, Giulio and Frellesvig, Hjalte and Pokraka, Andrzej and Smith, Sid",
    title = "{Magic Relations and Critical Varieties of Feynman Integrals}",
    eprint = "2605.29789",
    archivePrefix = "arXiv",
    primaryClass = "hep-th",
    month = "5",
    year = "2026"
}

@inproceedings{Britto:2024mna,
    author = "Britto, Ruth and Duhr, Claude and Hannesdottir, Holmfridur S. and Mizera, Sebastian",
    title = "{Cutting-Edge Tools for Cutting Edges}",
    eprint = "2402.19415",
    archivePrefix = "arXiv",
    primaryClass = "hep-th",
    reportNumber = "BONN-TH-2024-05",
    doi = "10.1016/B978-0-323-95703-8.00097-5",
    month = "2",
    year = "2024"
}

@article{Lee:2010ik,
    author = "Lee, R. N. and Smirnov, V. A.",
    title = "{Analytic Epsilon Expansions of Master Integrals Corresponding to Massless Three-Loop Form Factors and Three-Loop g-2 up to Four-Loop Transcendentality Weight}",
    eprint = "1010.1334",
    archivePrefix = "arXiv",
    primaryClass = "hep-ph",
    doi = "10.1007/JHEP02(2011)102",
    journal = "JHEP",
    volume = "02",
    pages = "102",
    year = "2011"
}

\end{document}